\documentclass[pra,reprint,superscriptaddress,longbibliography]{revtex4-2}

\usepackage{array,caption}
\usepackage{natbib}
\usepackage[section]{placeins}
\usepackage{pgfplots,amsfonts}
\usepackage[breaklinks]{hyperref}
\usepackage{amssymb}
\usepackage{amsmath,amsthm,bm}
\usepackage{amsfonts}
\usepackage{cleveref}
\usepackage{subcaption}
\usepackage{mathtools}
\usepackage{tikz}
\hypersetup{citecolor=red,colorlinks=true,urlcolor=blue}
\usepackage{algorithm,setspace}
\usepackage{algpseudocode}
\usetikzlibrary{angles, quotes}
\usetikzlibrary{positioning}

\usepackage[normalem]{ulem}
\usepackage{color}
\usepackage{xcolor}

\newcolumntype{P}[1]{>{\centering\arraybackslash}p{#1}}

\newcommand{\ket}[1]{\mbox{$| #1 \rangle$}}

\definecolor{Ugreen}{HTML}{198a11}

\begin{document}
\title{Characterizing conical intersections of nucleobases on quantum computers}

\author{Yuchen Wang}
\affiliation{Department of Chemistry and The James Franck Institute, The University of Chicago, Chicago, Illinois 60637, USA}
\author{Cameron Cianci}
\affiliation{Department of Physics, University of Connecticut, Storrs, Connecticut 06269, USA}
\affiliation{Mirion Technologies (Canberra) Inc., 800 Research Parkway, Meriden, Connecticut 06450, USA}
\author{Irma Avdic}
\affiliation{Department of Chemistry and The James Franck Institute, The University of Chicago, Chicago, Illinois 60637, USA}
\author{Rishab Dutta}
\affiliation{Department of Chemistry, Yale University, P.O. Box 208107, New Haven, Connecticut 06520-8107, USA}
\author{Samuel Warren}
\affiliation{Department of Chemistry and The James Franck Institute, The University of Chicago, Chicago, Illinois 60637, USA}
\author{Brandon Allen}
\affiliation{Department of Chemistry, Yale University, P.O. Box 208107, New Haven, Connecticut 06520-8107, USA}
\author{Nam P. Vu}
\affiliation{Department of Chemistry, Yale University, P.O. Box 208107, New Haven, Connecticut 06520-8107, USA}
\author{Lea F. Santos}
\affiliation{Department of Physics, University of Connecticut, Storrs, Connecticut 06269, USA}
\author{Victor S. Batista}
\affiliation{Department of Chemistry, Yale University, P.O. Box 208107, New Haven, Connecticut 06520-8107, USA}
\author{David A. Mazziotti}
\affiliation{Department of Chemistry and The James Franck Institute, The University of Chicago, Chicago, Illinois 60637, USA}

\date{Submitted October 24, 2024}

\begin{abstract}
Hybrid quantum-classical computing algorithms offer significant potential for accelerating the calculation of the electronic structure of strongly correlated molecules. In this work, we present the first quantum simulation of conical intersections (CIs) in a biomolecule, cytosine, using a superconducting quantum computer. We apply the Contracted Quantum Eigensolver (CQE)---with comparisons to conventional Variational Quantum Deflation (VQD)---to compute the near-degenerate ground and excited states associated with the conical intersection, a key feature governing the photostability of DNA and RNA. 
The CQE is based on an exact ansatz for many-electron molecules in the absence of noise---a critically important property for resolving strongly correlated states at CIs.  Both methods demonstrate promising accuracy when compared with exact diagonalization, even on noisy intermediate-scale quantum computers, highlighting their potential for advancing the understanding of photochemical and photobiological processes. The ability to simulate these intersections is critical for advancing our knowledge of biological processes like DNA repair and mutation, with potential implications for molecular biology and medical research.
\end{abstract}

\maketitle

\section{Introduction}

Recent advances in quantum computing open new horizons for the field of quantum chemistry with particularly promising applications in electronic structure simulations~\cite{aspuru2005,cao2019,bauer2020,dutta2024}. Various quantum algorithms that demonstrate a potential complexity advantage over classical ones have been proposed. Though still in the noisy intermediate-scale quantum (NISQ) era~\cite{preskill2018}, electronic structure simulations on quantum devices with appropriate error mitigation are poised to catch up to existing classical methods~\cite{kandala2017, russo2021, tang2021, sureshbabu2021, smart2022benzyne, dutta20242, cianci2024subspacesearchquantumimaginarytime}. An apparent application of quantum computing to biology is the electronic structure computation of moderately sized biomolecules~\cite{Cao2018, smaldone2024}. For these molecules, most classical wavefunction-based methods are restricted by a high computational scaling with the number of electrons. While resource-saving methods such as density functional theory can give a reasonable description of excited states~\cite{varsano2006, teh2019}, they lack the multireference character to describe degenerate electronic states, which are crucial for understanding many photochemical and photobiological processes.

This work explores the nonadiabatic decay of a nucleobase, cytosine, in its photo-excited state. As key components of deoxyribonucleic acid (DNA) and ribonucleic acid (RNA), nucleobases are intrinsically stable when exposed to ultraviolet radiation, preventing the mutations and genetic instability of DNA and RNA, such as those found in many forms of skin cancer~\cite{marrot2008}. Numerous studies have been conducted on the photobehavior of nucleobases, both from experimental and theoretical perspectives~\cite{kang2002, malone2003, sharonov2003, merchan2003, blancafort2004, matsika2005,kistler2008, gonzalez2010, barbatti2011, richter2012, blaser2016, trachsel2020, shahrokh2021}. Experimental observations reveal that photo-excited nucleobases exhibit remarkably short lifetimes and low fluorescence yields. These findings suggest that the excited molecules undergo rapid internal conversion, returning to their ground states via conical intersections (CIs) on ultrafast timescales.

CIs are the subspace formed by molecular geometries at which the electronic states are degenerate in energy~\cite{yarkony1996, domcke2004, levine2007, matsika2011, tully2012, guo2016, yarkony2019}. They are known to be the essential pathway for molecules going through internal conversion, ``funneling'' the population from excited states to the ground state. The first step toward simulating nonadiabatic molecular dynamics involving multiple electronic states is to characterize the CIs. The degeneracy in energy results in highly multireference character of adiabatic wavefunctions, which makes single-referenced methods no longer feasible. Additionally, since CIs are seams embedded in the potential energy manifold, they require electronic structure calculations involving batches of molecular geometries, which is a time-consuming task. Quantum computers offer the potential to accelerate electronic structure calculations. Consequently, the development of excited-state quantum algorithms capable of accurately describing CIs on NISQ devices is a critical prerequisite for future research in nonadiabatic molecular dynamics using quantum computing platforms~\cite{yalouz2021,ollitrault2020,wang2024,koridon2024,zhao2024}.

This study presents to our knowledge the first quantum simulation of CIs in a biomolecule, marking an initial step towards integrating nonadiabatic dynamics into quantum computing platforms and exploring the application of quantum computing in chemical biology. We introduce and test two well-documented quantum algorithms to compute the ground and excited states. The first one is the variational quantum deflation (VQD) and the second one is the contracted quantum eigensolver (CQE). Unlike ref.~\cite{zhao2024}, which simulated CIs by computing the ground states for two separate symmetry blocks, these algorithms are designed to compute multiple states within the same symmetry block, suitable for the treatment of accidental CIs~\cite{yarkony2019}. This paper is organized as follows: in the Theory section, we introduce the two quantum algorithms and then provide a brief review of the theory surrounding CIs; in the Results section, we present quantum simulations on noiseless and noisy fake backends, as well as on a 127-qubit IBM quantum computer; finally, we provide conclusions and outlook.

\section{Theory}

\subsection{Variational Quantum Deflation}

The variational quantum deflation (VQD) algorithm is an algorithm to find the $k$-lowest eigenvalues of a matrix~\cite{higgott2019}. VQD computes excited states by introducing a deflation term accounting for the orthogonality of eigenvectors. For a system described by the many-electron Schr\"{o}dinger equation (SE),
\begin{equation}
    (\hat{H}-E) | \Psi \rangle = 0,
\end{equation}
the cost function being optimized takes the following form,
\begin{equation}
    J({\theta_k}) 
    = \langle \Psi(\theta_k) | \hat{H} | \Psi(\theta_k) \rangle 
    + \sum_{j = 0}^{k-1} \beta | \langle  \Psi(\theta_j) | \Psi(\theta_k) \rangle | ^2 ,
\end{equation}
in which the $k$-th wavefunction is represented by a parameterized ansatz $| \Psi(\theta_k) \rangle$ with parameter set $\{ \theta_k \}$ and $\beta$ is the weight for the non-orthogonal penalty function. 
The algorithm can be understood as a constrained search that finds the energy minima of the $k$-th state subject to the constraints that the wavefunction of the target state must be orthogonal to all $(k-1)$ lower states. In excited state algorithms, the orthogonality condition is important to prevent the wavefunction from collapsing into a lower state during optimization. With an appropriate $\beta$ value, VQD achieves energy minimization while retaining the orthogonality of wavefunctions, proving to be an efficient and widely-used excited state algorithm on NISQ devices.

\subsection{Contracted Quantum Eigensolver}

The contracted quantum eigensolver (CQE)~\cite{smart2021,smart2022,wang2023CQE,warren2024} is a quantum algorithm originating from the contracted Schr\"{o}dinger equation (CSE)~\cite{Mazziotti1998,Mazziotti2006}, which contracts the SE onto the space of two electrons. Since a molecular Hamiltonian contains up to two-body interaction, the contraction is lossless in the sense that the CSE and the SE share an equivalent set of pure-state solutions~\cite{Mazziotti1998, Nakatsuji1976}.

We have introduced two modifications for CQE to tackle excited states. One is through performing variation on an ensemble (subspace) composed of orthogonal pure states~\cite{benavides2024} and the other is by replacing the Hamiltonian with the variance~\cite{wang2023}. The latter is used in this work as it is a state-specific method that can converge to stationary solutions regardless of the energy gap. We first briefly review the method and then provide some comments on different modifications and their performance in characterizing CIs.

The variance of SE is defined as 
\begin{equation}\label{eq:var}
    {\rm Var} = \langle \Psi | ( {\hat H} – E )^{2} | \Psi \rangle.
\end{equation}
The variance vanishes only when the SE converges to a stationary solution, which in turn allows us to use the non-vanishing residual of the variance to construct an exponential transformation to update the non-stationary wavefunction. 

\begin{figure}[htbp]
    \includegraphics[width=0.5\textwidth]{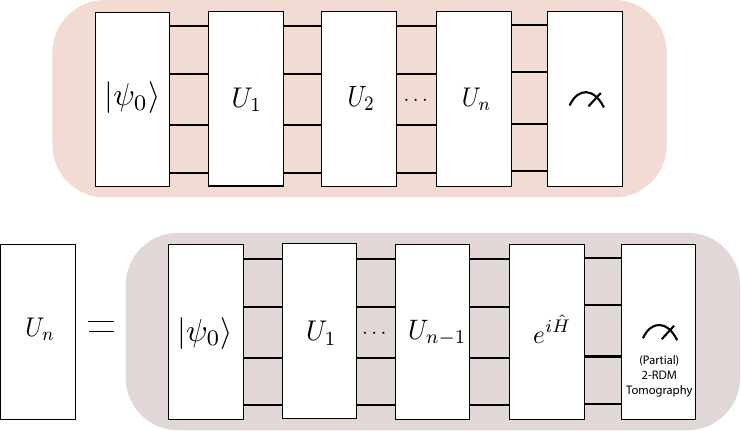}
    \caption{An illustration of the CQE algorithm at the $n$th iteration.}
    \label{fig:schematic}
\end{figure}

In variance-based CQE, we minimize the variance at the $m$-th iteration with respect to the two-body anti-Hermitian operator $\hat F_{m}$,
\begin{equation}
{\hat F}_{m} = \sum_{pqst}{ ^{2} F^{pq;st}_{m} {\hat a}^{\dagger}_{p} {\hat a}^{\dagger}_{q} {\hat a}^{}_{t} {\hat a}^{}_{s} },
\end{equation}
and the wavefunction ansatz is a product of the residuals in Eq.~\ref{eq:var},
\begin{equation}
    \ket{\Psi_{m}} = \prod^m e^{\epsilon\hat{F}_{m}}\ket{\Psi_0}.
\end{equation}
Here ${\hat a}^{\dagger}_{i}$ and ${\hat a}_{i}$ are the creation and annihilation operators with respect to the $i$-th orbital.  The parameter $\epsilon$ is the learning rate that can be optimized for better convergence. In each iteration, $\hat{F}$ is evaluated with quantum state tomography. An illustration of the CQE algorithm is given in Fig.~\ref{fig:schematic}. The algorithm can be characterized as an iterative update of the wavefunction based on the ``on-the-fly" residual of the CSE.

One difficulty of using the energy variance rather than energy itself is that squaring the Hamiltonian operators typically introduces additional terms that increase the measurement cost . We have used a second-order Taylor expansion to approximate the variance and related variables by preparing an auxiliary state as follows~\cite{wang2023},
\begin{equation}
\label{eq:tilde}
| {\tilde \Psi_{m}} \rangle = {\rm e}^{i \delta \left ( {\hat H} - E_{m} \right ) } | \Psi_{m} \rangle,
\end{equation}
and then measuring the variance using the following equation,
\begin{equation}
\label{eq:mvar}
\langle \Psi_{m} | ( {\hat H} - E_{m} )^{2} | \Psi_{m} \rangle \approx \frac{ 1 - \Re \langle \Psi_{m} | {\tilde \Psi}_{m} \rangle}{\delta^{2}/2},
\end{equation}
where the real part of the overlap can be estimated with a Hadamard test procedure~\cite{cleve1998}, reducing the measurement cost with an ancillary qubit. 
We present the iterative algorithm of variance-based CQE in table~\ref{tab:alg}.

\begin{table}[h]
  \caption{\normalsize Variance-based CQE algorithm.}
  \label{tab:alg}
  \begin{ruledtabular}
  \begin{tabular}{l}
  {\bf Algorithm: Variance-based CQE} \\
  \hspace{0.0in} {\rm Given} $m=0$ {\rm and convergence tolerance} $\epsilon$. \\
  \hspace{0.0in} {\rm Choose initial wave function} $| \Psi_{0} \rangle$. \\
  \hspace{0.0in} {\rm Repeat until the energy variance is less than} $\epsilon$. \\
  \hspace{0.1in} {\rm {\bf Step 1:} Prepare} $| \tilde \Psi_{m} \rangle = {\rm e}^{i \delta ( {\hat H} - E_{m} )} | \Psi_{m} \rangle$  \\
  \hspace{0.1in} {\rm {\bf Step 2:} Measure} $F^{pq;st}_{m+1}$ and compute ${\hat F}_{m+1}$ \\
  \hspace{0.1in} {\rm {\bf Step 3:} Prepare} $| \Psi_{m+1} \rangle = {\rm e}^{\epsilon {\hat F}_{m+1}} | \Psi_{m} \rangle$  \\
  \hspace{0.1in} {\rm {\bf Step 4:} Optimize variance with respect to} $\epsilon$ \\
  \hspace{0.1in} {\rm {\bf Step 5:} Set} $m = m+1$.
  \end{tabular}
  \end{ruledtabular}
\end{table}

By using the energy variance rather than the energy as the optimization target, the variance-based CQE ensures convergence to local minima, significantly enhancing numerical stability~\cite{wang2023,alcoba2024}. Moreover, it does not require information about lower-energy states, enabling the targeting of specific states during slow-varying changes in molecular geometry. This feature makes the algorithm particularly suitable for describing CIs. The disadvantage of this approach is that orthogonality between wavefunctions is not guaranteed, which means the target CI state could converge to the other energetically degenerate eigenstate during the optimization process. To address this concern, we use an additional orthogonality check procedure, which involves calculating the overlap between the two converged state vectors. Our simulations did not reveal any instances of such collapsing behavior.

Other popular modifications for calculating the $k$th excited states, such as the deflation term in VQD and the subspace modification found in SSVQE~\cite{nakanishi2019} and parallel CQE~\cite{benavides2024}, all guarantee the orthogonality of adiabatic wavefunctions during optimization. However, they require converging all low-lying states either in advance or simultaneously, which can become hard when $k$ is large. Another potential issue is related to the convergence for CIs. At the molecular geometry where energies are exactly degenerate, two eigenvectors can be linearly combined to form new and non-orthogonal eigenvectors~\cite{yarkony1996}. We expect this to slow down the convergence of excited state algorithms based on orthogonality. 

\subsection{Conical Intersections}

In this section, we provide a brief review of key concepts regarding CIs. The diabatic electronic Schr\"{o}dinger equation is
\begin{equation}\label{dse}
    [\mathbf{H^{d}}(\mathbf{R}) - \mathbf{I}E_{J}(\mathbf{R})]\mathbf{d}^{J}(\mathbf{R})= \mathbf{0},
\end{equation}
where $\mathbf{H^{d}}$ is the diabatic Hamiltonian matrix with dimension $(N^{state}, N^{state})$, and $E_{J}$ is the energy of the $J$th state. While CIs between more than two states are possible, we consider $N^{state}=2$ here with the analysis being generalizable. The CIs between two electronic states form only when the two following constraints are simultaneously satisfied,
\begin{equation}\label{eq:CI}
H_{11}^{d}=H_{22}^{d}, H_{12}^{d}=H_{21}^{d}=0,
\end{equation}
where $H_{IJ}^d$ is the matrix element of $\mathbf{H^{d}}$. Eq.~\ref{eq:CI} imposed on the global potential energy matrices gives us the $(N-2)$ dimensional seam of CIs. We define the $\mathbf{g}$, $\mathbf{h}$ vectors that lift the branching space~\cite{yarkony1996},
\begin{equation}\label{eq:gh}
    \mathbf{g} = \frac{1}{2}\nabla_{\mathbf{R}}(H_{11}-H_{22}), \mathbf{h} = \nabla_{\mathbf{R}} H_{12}.
\end{equation}
The $\mathbf{g}$-$\mathbf{h}$ plane is important in understanding nonadiabatic dynamics because the molecules break their degeneracy and achieve a state transition only when moving within the plane, which provides insights on the mode-selective dynamics of nonadiabatic reactions.

The procedure to locate the CI seam and the minimum energy CI (MECI) point in this work is reported in previous literature~\cite{wang2024} and a classical implementation can be found in COLUMBUS~\cite{lischka2011,lischka2020,manaa1993}. We use a constrained Lagrangian defined below,
\begin{equation}
L(\mathbf{R})=E_{I}(\mathbf{R})+\lambda_{0}(E_{1}-E_{2})+\sum_{k=1}^{M} \lambda_{k}C_{k}(\mathbf{R}),
\label{eq:optimize}
\end{equation}
where $C_{k}$ are geometric constraints. The Lagrangian is minimized with a Newton-Raphson procedure to find the energy minima subject to energy degeneracy and additional geometry constraints. The gradient needed for the optimization can be obtained classically through analytical gradient techniques or on quantum computers with a finite difference method.



\section{Results}

\subsection{Classical Calculation Results}

The electronic structure calculations are performed with state-averaged complete active space self consistent field (SA-CASSCF) in COLUMBUS~\cite{lischka2011,lischka2020}. The first two singlets are averaged with equal weights. We use the correlation-consistent polarized valence double-zeta (cc-pVDZ) basis set and an active space of four electrons in three orbitals. Note that the choice of active space is limited by the current performance of NISQ devices rather than the quantum algorithms. In our analysis, energy levels are considered degenerate if their difference is less than 0.0005~hartree ($\sim$100 cm$^{-1}$) as determined by exact diagonalization. This level of precision is deemed sufficient given the choice of active space and basis set. The Hamiltonian is constructed from electron integrals in the CASSCF orbital basis. As the Hamiltonian does not guarantee the spin multiplicity ($S^2$) of the wavefunction, we perform an additional check to ensure the $S^2$ values correspond to singlet states.

The CI in cytosine is found along several active vibrational modes---a situation that contributes to the complexity of its rich photochemical behavior. We only investigate the CIs between the first two singlets, which excludes the three-state CIs reported in several previous works~\cite{blancafort2004,matsika2005,kistler2008,gonzalez2010}. There are multiple CIs for the first two singlet states of cytosine~\cite{cuellar2023}. Here we focus on the $\pi\pi^{*}/S_{0}$ CI, which has been characterized as the major reaction intermediate for its internal conversion.

\begin{figure}[htbp]
    \includegraphics[width=0.4\textwidth]{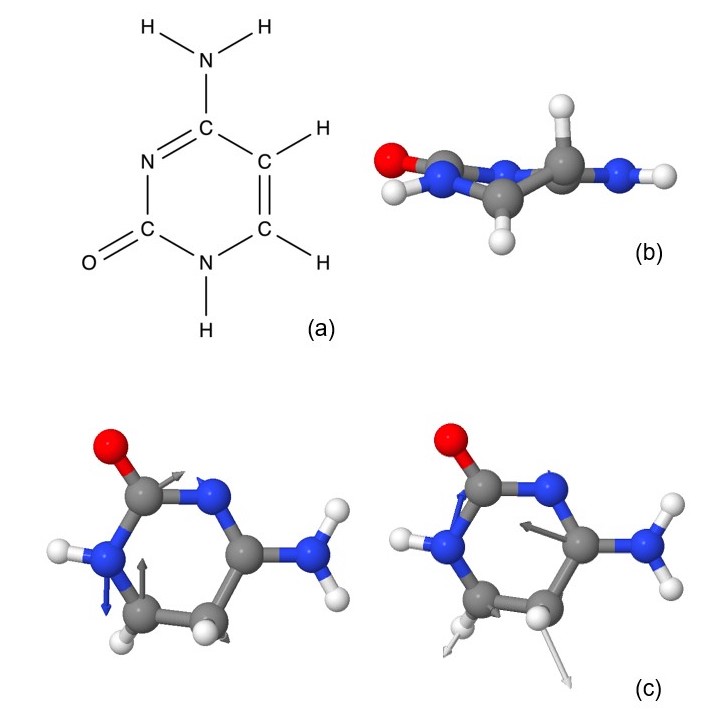}
    \caption{Reported minimum energy CI for cytosine. (a) chemical structure, (b) planar perspective of the ring structure, (c) the orthogonalized $\mathbf{g}$ (left), $\mathbf{h}$ (right) vectors at CI}
    \label{fig:classical}
\end{figure}

Figure~\ref{fig:classical} reports the minimum energy point on the $\pi\pi^{*}/S_{0}$ CIs. That point is optimized classically to an energy difference between the two CI states of less than 0.0005 and a norm of the constrained Lagrangian in Eq.~(\ref{eq:optimize}) of less than 0.01. The CI between $\pi\pi^{*}$ and $S_{0}$ occurs at a distorted molecular structure. The vibrational mode leading to the broken degeneracy is assigned to the out-of-plane torsion (both $\mathbf{g}$ and $\mathbf{h}$) as well as the vibration of the non-coplanar hydrogen ($\mathbf{h}$).




\subsection{Quantum Simulation Results}

While classical calculations capture the electronic structure information of the system within the limits of the employed active space, we now examine the potential applications of quantum computers in characterizing CIs. Two key issues are addressed: first, whether a given quantum algorithm can describe the energy degeneracy and potential energy surface topography in the vicinity of CIs and second, whether CIs can be efficiently optimized with a hybrid method that employs classical optimization and quantum simulation. We have used three different backends in Qiskit~\cite{Qiskit} in this work:

(i) an ideal statevector simulator without noise, which we use to verify the exactness of the tested algorithms and provide an estimation of the convergence speed in a noiseless environment. For an ideal simulator, the algorithms are not limited by the number of qubits. We thus use the Jordan-Wigner mapping~\cite{Jordan1928}, which is generally a sparse mapping, to map the Hamiltonian to six qubits.

(ii) a fake backend FakeSherbrooke, which we use to mimic the behavior of IBM Sherbrooke~\cite{ibm_quantum}.  The Fake backend provides a playground for classical optimizers that may require a significant amount of quantum resources. The number of qubits has a significant effect on the quantum algorithm performance in the presence of noise. We employ additional tapering techniques, based on conserving of $N$ (number of electrons) and $S_z$ (total spin number), to reduce the qubits required to four, which is the minimum number of qubits required to avoid truncating the Hamiltonian matrix. An implementation can be found in the ParityMapper in Qiskit~\cite{Qiskit}.

(iii) The 127-qubit IBM Cleveland with the Eagle r3 type processor~\cite{ibm_quantum}, which we use to perform experimental simulations.  While IBM Cleveland and Sherbrooke share the same generation of quantum processors, their noise behavior can be quite different.

\begin{figure}[h]
    \begin{subfigure}[b]{0.5\textwidth}
        \includegraphics[width=0.8\textwidth]{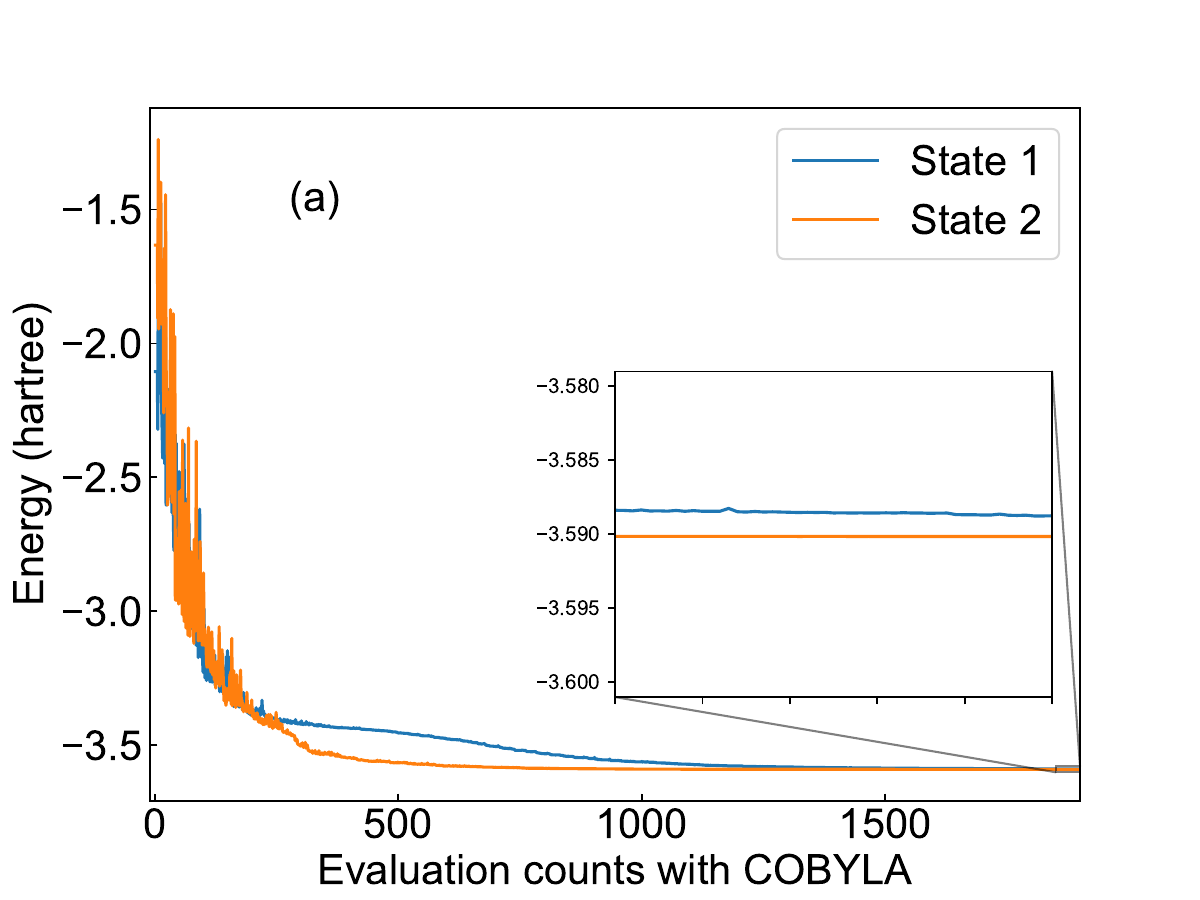}
    \end{subfigure}
    \begin{subfigure}[b]{0.5\textwidth}
        \includegraphics[width=0.8\textwidth]{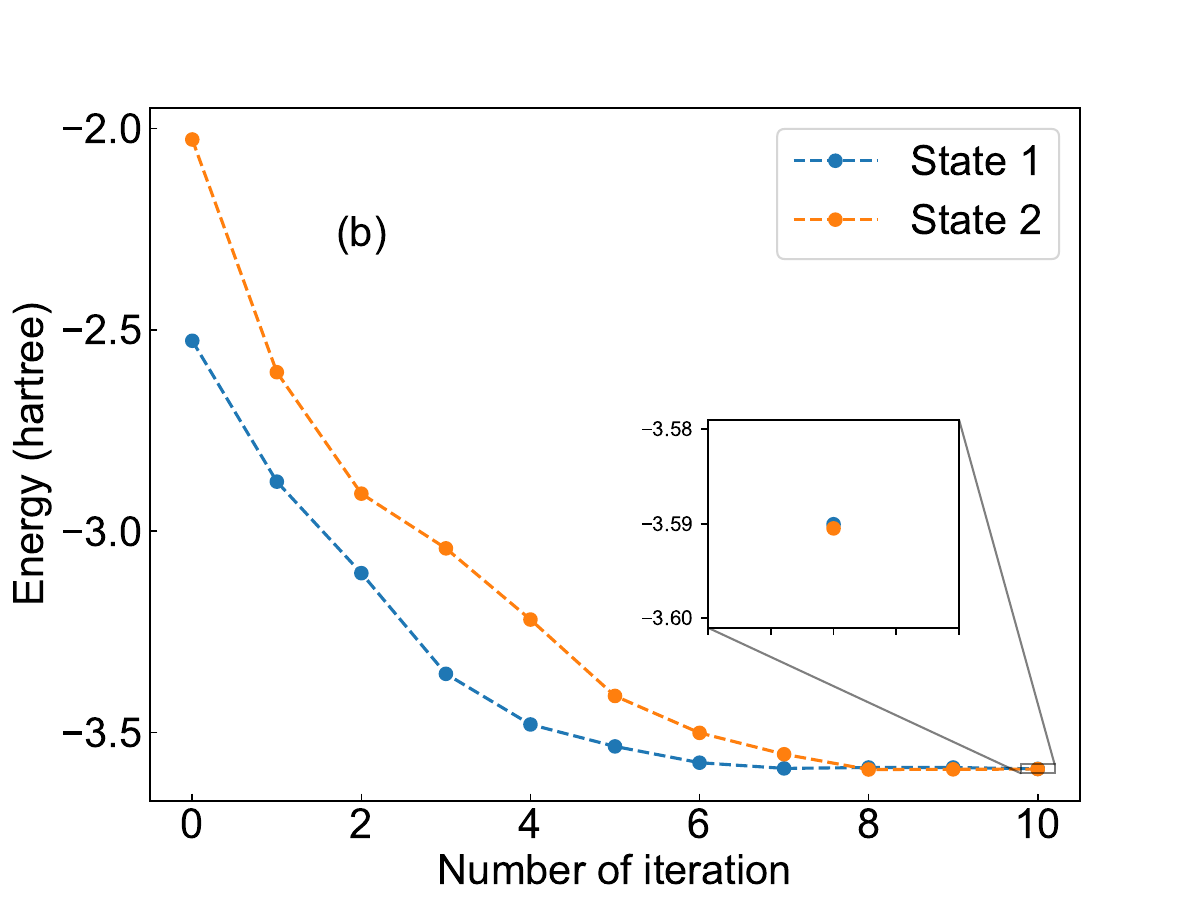}
    \end{subfigure}
    \caption{The convergence plot for (a) VQD and (b) CQE on a noiseless simulator. Both ground and first excited singlets are shown. For the energy values on the y axis, the core energy has been excluded from the total energy. }
    \label{fig:noiseless}
\end{figure}

We report the convergence of both algorithms for a CI Hamiltonian on a noiseless statevector simulator in Fig.~\ref{fig:noiseless}. For the VQD calculation, we employ the implementation in Qiskit Algorithm library. A two-local ansatz with four layers is used, which gives a circuit depth of 19. A COBYLA optimizer in Qiskit is used for the classical variation. Figure~\ref{fig:noiseless} (a) shows that VQD converges to the threshold with a sufficiently large number of evaluations. The two-local ansatz used here is adequate to parameterize the near-degenerate ground- and excited-state wavefunctions to the desired accuracy in noiseless simulations.

CQE has been demonstrated as an exact ansatz~\cite{smart2024} and in the noiseless environment we do not restrict the number of unitaries being applied to verify the exactness. The two states are optimized separately with single Slater determinants as the initial guesses. For the noiseless statevector simulation, it is shown in Fig.~\ref{fig:noiseless}(b) that CQE converges to the exact solution in fewer than 10 iterations, which is noteworthy given the quantum resources.

\begin{figure}[h]
    \begin{subfigure}[b]{0.5\textwidth}
        \includegraphics[width=0.8\textwidth]{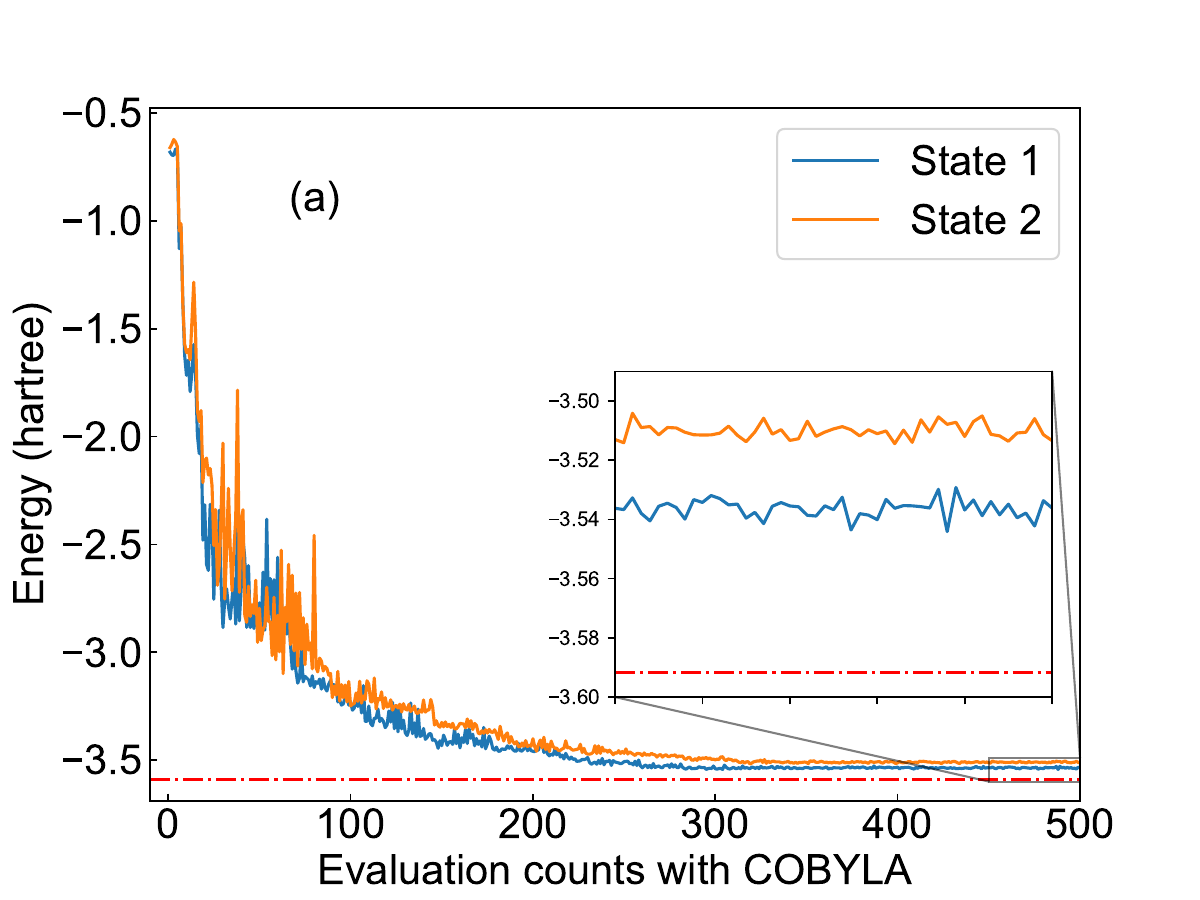}
    \end{subfigure}
    \begin{subfigure}[b]{0.5\textwidth}
        \includegraphics[width=0.8\textwidth]{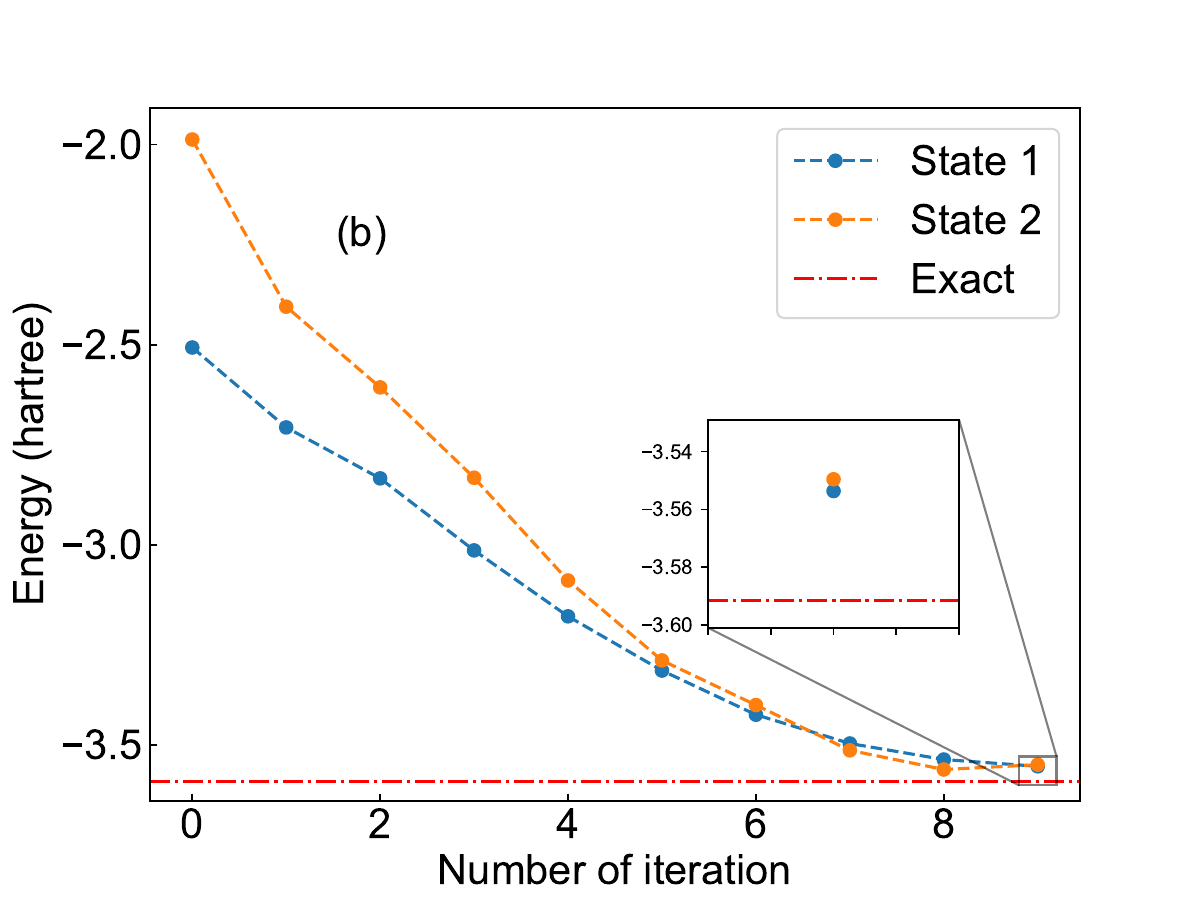}
    \end{subfigure}
    \caption{The convergence plot for (a) VQD and (b) CQE on FakeSherbrooke. Both ground and first excited singlets are shown. For the energy values on the y axis, the core energy has been excluded from the total energy.}
    \label{fig:fake_backend}
\end{figure}

We next examine the convergence on the FakeSherbrooke backend. The circuit depth has a significant impact on the algorithm performance in the presence of noise. In order to make a fair comparison, we have restricted the number of iterations and unitaries per iteration of CQE to generate an ansatz with almost the same circuit depth as VQD. We apply 8192~shots to every circuit. The exact value from diagonalization is plotted for comparison. For VQD and CQE, the absolute errors without any error mitigation techniques are $\sim$60 mhartree and $\sim$40 mhartree, respectively, which is a consequence of the quantum noise. While the absolute error could be improved with error mitigation techniques, we here focus only on the description of the energy gap from the unmitigated results. As can be seen from Fig.~\ref{fig:fake_backend}, the error of the energy gap is only $\sim$20 mhartree for VQD and single-digit mhartree at the ninth iteration of CQE. In both methods the energy gap exhibits a smaller error than the absolute energies due to error cancellation. 

We observe that CQE outperforms VQD in characterizing the CIs.  One reason, discussed in the Theory section, is that the variance-based CQE is a state-specific method, while VQD is dependent upon the given state and any energetically equal or lower states, which can slow the convergence for near-degenerate states. Another possible reason is that the CQE uses the ansatz from the ACSE~\cite{Mazziotti.2007, Mazziotti.2004}, which is a physics-informed exact ansatz, while the VQD, implemented in this work, employs a two-local ansatz that does not account for the structure of the Hamiltonian or other physical insights into the system. A more efficient ansatz, such as the unitary coupled cluster~\cite{Hoffmann1988,romero2018} or the ACSE ansatz~\cite{Smart.2021, warren2024, Mazziotti.2007, Mazziotti.2004}, is advised for VQD to achieve better performance.

One may also draw a rough comparison between the quantum resources. It takes around 320 evaluations for VQD to converge with the COBYLA optimizer, while it takes eight iterations for CQE. We note that for each iteration, the CQE requires a two-particle reduced density matrix (2-RDM) partial tomography procedure, leading to a similar number of measurements. However, the quantum state tomography can be implemented in a parallel way, favoring the CQE in time complexity. Moreover, because the CQE exploits the restriction to pairwise interactions in the Hamiltonian, the CQE for any number $N$ of electrons depends only on the 2-RDM, which has a scaling of $O(r^4)$ with orbital number $r$ (with tomography protocol even less~\cite{bonet2020}), making it a good candidate for scalable molecular simulations with large active spaces.

\begin{table*}[t]
\caption{\centering Optimization process for CIs on quantum computers. Energies are reported with respect to the global minima of -392.6536. Unit is hartree.}
\centering
\begin{tabular}{|c|c|c|c|c|c|c|}
\hline
Iteration & E$_0$(simulation) & E$_1$(simulation) & $\Delta $E(simulation) & E$_0$(exact) & E$_1$(exact) & $\Delta $E(exact) \\  
\hline
0 & 0.1620 & 0.2313 & 0.0693 & 0.1401 & 0.2081 & 0.0680 \\
1 & 0.1714 & 0.2229 & 0.0515 & 0.1408 & 0.1964 & 0.0556 \\
2 & 0.1812 & 0.2112 & 0.0300 & 0.1416 & 0.1843 & 0.0427 \\
3 & 0.1897 & 0.2036 & 0.0139 & 0.1430 & 0.1684 & 0.0254 \\
4 & 0.1770 & 0.1959 & 0.0189 & 0.1436 & 0.1618 & 0.0182 \\
5 & 0.1762 & 0.1841 & 0.0079 & 0.1449 & 0.1551 & 0.0102 \\
\hline
\end{tabular}
\label{tab:fourcolumn}
\end{table*}

After the fake backend experiments, we use the CQE to characterize the topography of the CI on a real IBM quantum computer. Several error mitigation techniques are used to suppress the error on IBM Cleveland, which are described in detail in the supporting information of ref.~\cite{wang20242}. We utilize Qiskit's built-in zero noise extrapolation, gate twirling and Twirled Readout Error eXtinction (TREX) for error mitigation~\cite{Qiskit}. Moreover, in the CQE we have applied a dynamical threshold to the unitaries~\cite{smart2022benzyne} in which after decomposing the unitaries in the Pauli basis, we select only those with coefficients above a threshold that decreases dynamically with convergence.

\begin{figure}[h!]
    \begin{subfigure}[b]{0.5\textwidth}
        \includegraphics[width=0.62\textwidth]{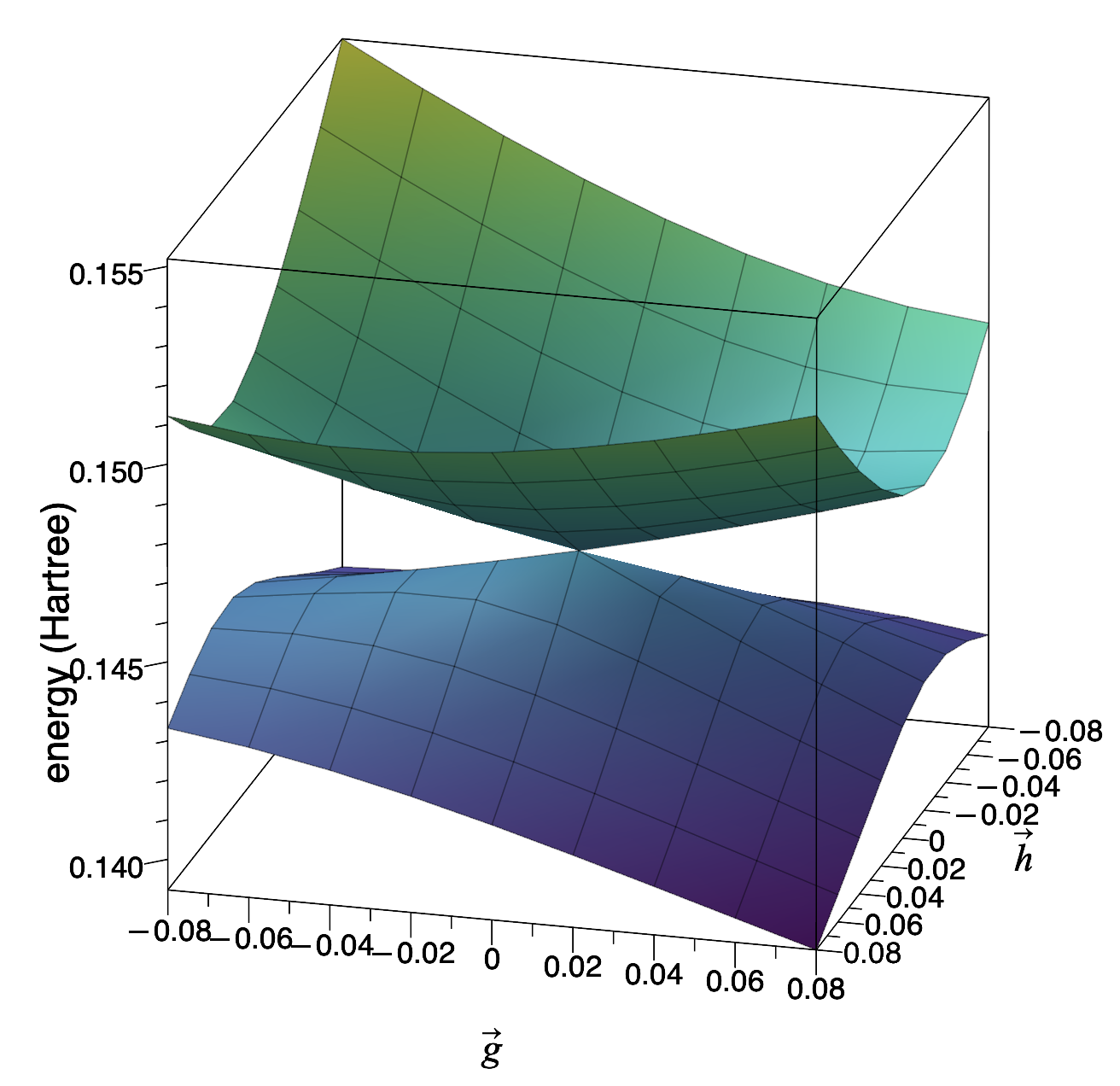}
    \end{subfigure}
    \begin{subfigure}[b]{0.5\textwidth}
        \includegraphics[width=0.6\textwidth]{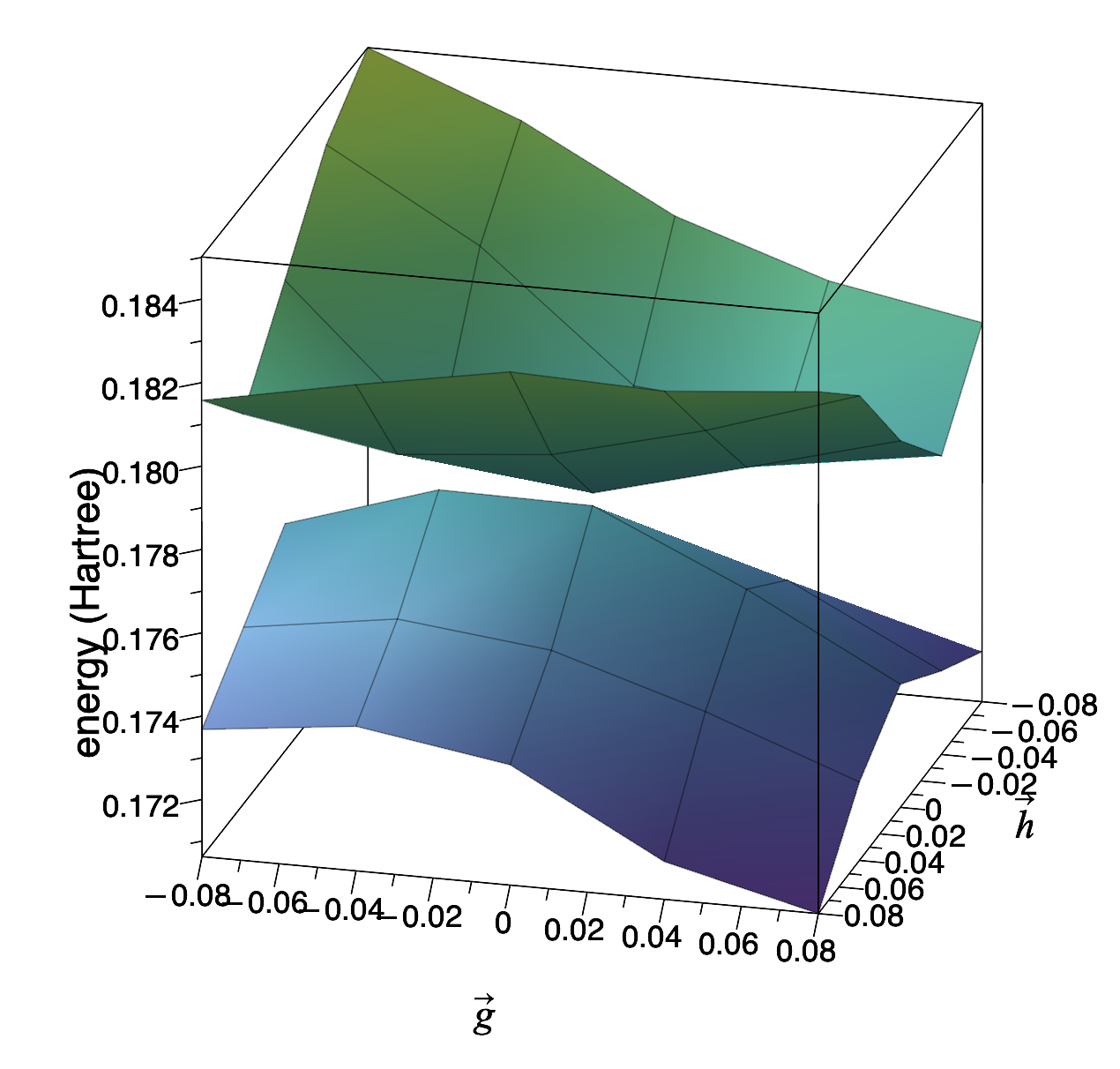}
    \end{subfigure}
    \caption{The 3D CI topography shown with (a) exact diagonalization (b) quantum simulation with IBM Cleveland. The energies in z axis are the relative values compared to global minima of -392.6536.}
    \label{fig:3d}
\end{figure}

We span the grid of molecular geometry along the $\mathbf{g}$, $\mathbf{h}$ directions and plot the coupled 3D potential energy surfaces in the vicinity of the CI. The exact surface and the surface from quantum simulation are shown in Fig.~\ref{fig:3d} (a) and (b),respectively. The surfaces from real quantum computers deviate slightly from the exact surfaces in absolute energies, and they are not as smooth as the exact ones. However, the topography of the double cone is generally well-preserved. The degenerate point in Fig.~\ref{fig:3d}(a) is no longer degenerate on quantum computers (Fig.~\ref{fig:3d}(b)) due to noise, but the two points are not far away. As has been observed in single-point calculations, if the error generated from quantum noise is uniform, then the only effect of noise is to shift the absolute value of the upper and lower surfaces by a certain amount, which in principle should not affect the nonadiabatic dynamics significantly.  The ability to characterize the electronic structure of the strongly correlated states at CIs provides an important step towards practical applications for the simulation of nonadiabatic dynamics on NISQ devices.

We further can locate CIs with a hybrid quantum-classical optimization method to minimize the energy gap as a function of the nuclear coordinates. Here we obtained numerical nuclear gradients by performing finite differences in Cartesian coordinates. The stepsize employed is 0.01. For the ease of implementation, we only allow the coordinate of the hydrogen atom with the greatest vibration amplitude in the $\mathbf{g}$-$\mathbf{h}$ plane to vary while keeping the rest of the molecule fixed. The optimization results are provided in Table~\ref{tab:fourcolumn}. At each iteration, we report the quantum simulation and exact diagonalization results at the same molecular geometry. The energy gap decreases at every iteration with the energy gap's error being well below the absolute error of individual states.  We are able to converge the geometry optimization within single mhartree accuracy, which is noteworthy given the noisy nature of current quantum computers. We also observe that the energy difference during optimization can be quite stochastic (e.g., between iterations 3 and 4) even though we only have three degrees of freedom. The reason for this behavior is that in the presence of noise, the nuclear gradients obtained from finite differences may subject to large deviations. A stochastic optimizer such as Simultaneous Perturbation Stochastic Approximation~\cite{spall1998} may provide better convergence when the quantum resources are limited and the number of nuclear degrees of freedom is larger.

\emph{Conclusions}---In this work we simulate the electronic structure of the ground and excited states of cytosine with quantum computers, focusing on the conical intersection---the crucial region for nonadiabatic dynamics simulations. We examine two quantum algorithms for excited state calculations: Variational Quantum Deflation (VQD) and Contracted Quantum Eigensolver (CQE), for performing crucial tasks for nonadiabatic dynamics such as the calculation of near-degenerate energies and the optimization of CI geometries. The CQE displays slightly better performance with cytosine due to its state-specific modification and the physics-informed ansatz.  For larger molecules and active spaces, the CQE would outperform the two-local VQD and other non-physics-based ansatzes.  Moreover, the CQE is based on the CSE ansatz---an exact ansatz for many-electron molecules in the absence of noise---, which is critical for resolving strongly correlated states such as those involved in the conical intersection.  Future work will examine the use of the CQE algorithm within nonadiabatic dynamic simulations. The present work represents an important first step towards harnessing the potential of quantum computers to provide more accurate and efficient decsriptions of conical interactions for accelerated progress in both photochemistry and photobiology. 

\textit{Acknowledgments}---D.A.M gratefully acknowledges the U.S. National Science Foundation Grant No. CHE-2155082 and the U.S. Department of Energy, Office of Basic Energy Sciences, Grant DE-SC0019215. I.A. gratefully acknowledges the NSF Graduate Research Fellowship Program under Grant No. 2140001. V.S.B and L.F.S  acknowledge support from the National Science Foundation Engines Development Award: Advancing Quantum Technologies (CT) under Award Number 2302908, and partial support from the National Science Foundation Center for Quantum Dynamics on Modular Quantum Devices (CQD-MQD) under Award Number 2124511. The views expressed are those of the authors and do not reflect the official policy or position of IBM or the IBMQ team.
\bibliography{refs}

\begin{thebibliography}{70}%
\makeatletter
\providecommand \@ifxundefined [1]{%
 \@ifx{#1\undefined}
}%
\providecommand \@ifnum [1]{%
 \ifnum #1\expandafter \@firstoftwo
 \else \expandafter \@secondoftwo
 \fi
}%
\providecommand \@ifx [1]{%
 \ifx #1\expandafter \@firstoftwo
 \else \expandafter \@secondoftwo
 \fi
}%
\providecommand \natexlab [1]{#1}%
\providecommand \enquote  [1]{``#1''}%
\providecommand \bibnamefont  [1]{#1}%
\providecommand \bibfnamefont [1]{#1}%
\providecommand \citenamefont [1]{#1}%
\providecommand \href@noop [0]{\@secondoftwo}%
\providecommand \href [0]{\begingroup \@sanitize@url \@href}%
\providecommand \@href[1]{\@@startlink{#1}\@@href}%
\providecommand \@@href[1]{\endgroup#1\@@endlink}%
\providecommand \@sanitize@url [0]{\catcode `\\12\catcode `\$12\catcode
  `\&12\catcode `\#12\catcode `\^12\catcode `\_12\catcode `\%12\relax}%
\providecommand \@@startlink[1]{}%
\providecommand \@@endlink[0]{}%
\providecommand \url  [0]{\begingroup\@sanitize@url \@url }%
\providecommand \@url [1]{\endgroup\@href {#1}{\urlprefix }}%
\providecommand \urlprefix  [0]{URL }%
\providecommand \Eprint [0]{\href }%
\providecommand \doibase [0]{https://doi.org/}%
\providecommand \selectlanguage [0]{\@gobble}%
\providecommand \bibinfo  [0]{\@secondoftwo}%
\providecommand \bibfield  [0]{\@secondoftwo}%
\providecommand \translation [1]{[#1]}%
\providecommand \BibitemOpen [0]{}%
\providecommand \bibitemStop [0]{}%
\providecommand \bibitemNoStop [0]{.\EOS\space}%
\providecommand \EOS [0]{\spacefactor3000\relax}%
\providecommand \BibitemShut  [1]{\csname bibitem#1\endcsname}%
\let\auto@bib@innerbib\@empty
\bibitem [{\citenamefont {Aspuru-Guzik}\ \emph {et~al.}(2005)\citenamefont
  {Aspuru-Guzik}, \citenamefont {Dutoi}, \citenamefont {Love},\ and\
  \citenamefont {Head-Gordon}}]{aspuru2005}%
  \BibitemOpen
  \bibfield  {author} {\bibinfo {author} {\bibfnamefont {A.}~\bibnamefont
  {Aspuru-Guzik}}, \bibinfo {author} {\bibfnamefont {A.~D.}\ \bibnamefont
  {Dutoi}}, \bibinfo {author} {\bibfnamefont {P.~J.}\ \bibnamefont {Love}},\
  and\ \bibinfo {author} {\bibfnamefont {M.}~\bibnamefont {Head-Gordon}},\
  }\bibfield  {title} {\bibinfo {title} {Simulated quantum computation of
  molecular energies},\ }\href {https://doi.org/10.1126/science.1113479}
  {\bibfield  {journal} {\bibinfo  {journal} {Science}\ }\textbf {\bibinfo
  {volume} {309}},\ \bibinfo {pages} {1704} (\bibinfo {year}
  {2005})}\BibitemShut {NoStop}%
\bibitem [{\citenamefont {Cao}\ \emph {et~al.}(2019)\citenamefont {Cao},
  \citenamefont {Romero}, \citenamefont {Olson}, \citenamefont {Degroote},
  \citenamefont {Johnson}, \citenamefont {Kieferov{\'a}}, \citenamefont
  {Kivlichan}, \citenamefont {Menke}, \citenamefont {Peropadre}, \citenamefont
  {Sawaya} \emph {et~al.}}]{cao2019}%
  \BibitemOpen
  \bibfield  {author} {\bibinfo {author} {\bibfnamefont {Y.}~\bibnamefont
  {Cao}}, \bibinfo {author} {\bibfnamefont {J.}~\bibnamefont {Romero}},
  \bibinfo {author} {\bibfnamefont {J.~P.}\ \bibnamefont {Olson}}, \bibinfo
  {author} {\bibfnamefont {M.}~\bibnamefont {Degroote}}, \bibinfo {author}
  {\bibfnamefont {P.~D.}\ \bibnamefont {Johnson}}, \bibinfo {author}
  {\bibfnamefont {M.}~\bibnamefont {Kieferov{\'a}}}, \bibinfo {author}
  {\bibfnamefont {I.~D.}\ \bibnamefont {Kivlichan}}, \bibinfo {author}
  {\bibfnamefont {T.}~\bibnamefont {Menke}}, \bibinfo {author} {\bibfnamefont
  {B.}~\bibnamefont {Peropadre}}, \bibinfo {author} {\bibfnamefont {N.~P.}\
  \bibnamefont {Sawaya}}, \emph {et~al.},\ }\bibfield  {title} {\bibinfo
  {title} {Quantum chemistry in the age of quantum computing},\ }\href
  {https://doi.org/10.1021/acs.chemrev.8b00803} {\bibfield  {journal} {\bibinfo
   {journal} {Chem. Rev.}\ }\textbf {\bibinfo {volume} {119}},\ \bibinfo
  {pages} {10856} (\bibinfo {year} {2019})}\BibitemShut {NoStop}%
\bibitem [{\citenamefont {Bauer}\ \emph {et~al.}(2020)\citenamefont {Bauer},
  \citenamefont {Bravyi}, \citenamefont {Motta},\ and\ \citenamefont
  {Chan}}]{bauer2020}%
  \BibitemOpen
  \bibfield  {author} {\bibinfo {author} {\bibfnamefont {B.}~\bibnamefont
  {Bauer}}, \bibinfo {author} {\bibfnamefont {S.}~\bibnamefont {Bravyi}},
  \bibinfo {author} {\bibfnamefont {M.}~\bibnamefont {Motta}},\ and\ \bibinfo
  {author} {\bibfnamefont {G.~K.-L.}\ \bibnamefont {Chan}},\ }\bibfield
  {title} {\bibinfo {title} {Quantum algorithms for quantum chemistry and
  quantum materials science},\ }\href
  {https://doi.org/10.1021/acs.chemrev.9b00829} {\bibfield  {journal} {\bibinfo
   {journal} {Chem. Rev.}\ }\textbf {\bibinfo {volume} {120}},\ \bibinfo
  {pages} {12685} (\bibinfo {year} {2020})}\BibitemShut {NoStop}%
\bibitem [{\citenamefont {Dutta}\ \emph
  {et~al.}(2024{\natexlab{a}})\citenamefont {Dutta}, \citenamefont {Cabral},
  \citenamefont {Lyu}, \citenamefont {Vu}, \citenamefont {Wang}, \citenamefont
  {Allen}, \citenamefont {Dan}, \citenamefont {Corti{\~n}as}, \citenamefont
  {Khazaei}, \citenamefont {Smart} \emph {et~al.}}]{dutta2024}%
  \BibitemOpen
  \bibfield  {author} {\bibinfo {author} {\bibfnamefont {R.}~\bibnamefont
  {Dutta}}, \bibinfo {author} {\bibfnamefont {D.~G.}\ \bibnamefont {Cabral}},
  \bibinfo {author} {\bibfnamefont {N.}~\bibnamefont {Lyu}}, \bibinfo {author}
  {\bibfnamefont {N.~P.}\ \bibnamefont {Vu}}, \bibinfo {author} {\bibfnamefont
  {Y.}~\bibnamefont {Wang}}, \bibinfo {author} {\bibfnamefont {B.}~\bibnamefont
  {Allen}}, \bibinfo {author} {\bibfnamefont {X.}~\bibnamefont {Dan}}, \bibinfo
  {author} {\bibfnamefont {R.~G.}\ \bibnamefont {Corti{\~n}as}}, \bibinfo
  {author} {\bibfnamefont {P.}~\bibnamefont {Khazaei}}, \bibinfo {author}
  {\bibfnamefont {S.~E.}\ \bibnamefont {Smart}}, \emph {et~al.},\ }\bibfield
  {title} {\bibinfo {title} {Simulating chemistry on bosonic quantum devices},\
  }\bibfield  {journal} {\bibinfo  {journal} {arXiv preprint arXiv:2404.10214}\
  }\href {https://doi.org/10.48550/arXiv.2404.10214}
  {10.48550/arXiv.2404.10214} (\bibinfo {year}
  {2024}{\natexlab{a}})\BibitemShut {NoStop}%
\bibitem [{\citenamefont {Preskill}(2018)}]{preskill2018}%
  \BibitemOpen
  \bibfield  {author} {\bibinfo {author} {\bibfnamefont {J.}~\bibnamefont
  {Preskill}},\ }\bibfield  {title} {\bibinfo {title} {Quantum computing in the
  nisq era and beyond},\ }\href {https://doi.org/10.22331/q-2018-08-06-79}
  {\bibfield  {journal} {\bibinfo  {journal} {Quantum}\ }\textbf {\bibinfo
  {volume} {2}},\ \bibinfo {pages} {79} (\bibinfo {year} {2018})}\BibitemShut
  {NoStop}%
\bibitem [{\citenamefont {Kandala}\ \emph {et~al.}(2017)\citenamefont
  {Kandala}, \citenamefont {Mezzacapo}, \citenamefont {Temme}, \citenamefont
  {Takita}, \citenamefont {Brink}, \citenamefont {Chow},\ and\ \citenamefont
  {Gambetta}}]{kandala2017}%
  \BibitemOpen
  \bibfield  {author} {\bibinfo {author} {\bibfnamefont {A.}~\bibnamefont
  {Kandala}}, \bibinfo {author} {\bibfnamefont {A.}~\bibnamefont {Mezzacapo}},
  \bibinfo {author} {\bibfnamefont {K.}~\bibnamefont {Temme}}, \bibinfo
  {author} {\bibfnamefont {M.}~\bibnamefont {Takita}}, \bibinfo {author}
  {\bibfnamefont {M.}~\bibnamefont {Brink}}, \bibinfo {author} {\bibfnamefont
  {J.~M.}\ \bibnamefont {Chow}},\ and\ \bibinfo {author} {\bibfnamefont
  {J.~M.}\ \bibnamefont {Gambetta}},\ }\bibfield  {title} {\bibinfo {title}
  {Hardware-efficient variational quantum eigensolver for small molecules and
  quantum magnets},\ }\href {https://doi.org/10.1038/nature23879} {\bibfield
  {journal} {\bibinfo  {journal} {Nature}\ }\textbf {\bibinfo {volume} {549}},\
  \bibinfo {pages} {242} (\bibinfo {year} {2017})}\BibitemShut {NoStop}%
\bibitem [{\citenamefont {Russo}\ \emph {et~al.}(2021)\citenamefont {Russo},
  \citenamefont {Rudinger}, \citenamefont {Morrison},\ and\ \citenamefont
  {Baczewski}}]{russo2021}%
  \BibitemOpen
  \bibfield  {author} {\bibinfo {author} {\bibfnamefont {A.~E.}\ \bibnamefont
  {Russo}}, \bibinfo {author} {\bibfnamefont {K.~M.}\ \bibnamefont {Rudinger}},
  \bibinfo {author} {\bibfnamefont {B.~C.}\ \bibnamefont {Morrison}},\ and\
  \bibinfo {author} {\bibfnamefont {A.~D.}\ \bibnamefont {Baczewski}},\
  }\bibfield  {title} {\bibinfo {title} {Evaluating energy differences on a
  quantum computer with robust phase estimation},\ }\href
  {https://doi.org/10.1103/PhysRevLett.126.210501} {\bibfield  {journal}
  {\bibinfo  {journal} {Phys. Rev. Lett.}\ }\textbf {\bibinfo {volume} {126}},\
  \bibinfo {pages} {210501} (\bibinfo {year} {2021})}\BibitemShut {NoStop}%
\bibitem [{\citenamefont {Tang}\ \emph {et~al.}(2021)\citenamefont {Tang},
  \citenamefont {Shkolnikov}, \citenamefont {Barron}, \citenamefont {Grimsley},
  \citenamefont {Mayhall}, \citenamefont {Barnes},\ and\ \citenamefont
  {Economou}}]{tang2021}%
  \BibitemOpen
  \bibfield  {author} {\bibinfo {author} {\bibfnamefont {H.~L.}\ \bibnamefont
  {Tang}}, \bibinfo {author} {\bibfnamefont {V.}~\bibnamefont {Shkolnikov}},
  \bibinfo {author} {\bibfnamefont {G.~S.}\ \bibnamefont {Barron}}, \bibinfo
  {author} {\bibfnamefont {H.~R.}\ \bibnamefont {Grimsley}}, \bibinfo {author}
  {\bibfnamefont {N.~J.}\ \bibnamefont {Mayhall}}, \bibinfo {author}
  {\bibfnamefont {E.}~\bibnamefont {Barnes}},\ and\ \bibinfo {author}
  {\bibfnamefont {S.~E.}\ \bibnamefont {Economou}},\ }\bibfield  {title}
  {\bibinfo {title} {qubit-adapt-vqe: An adaptive algorithm for constructing
  hardware-efficient ans{\"a}tze on a quantum processor},\ }\href
  {https://doi.org/10.1103/PRXQuantum.2.020310} {\bibfield  {journal} {\bibinfo
   {journal} {PRX Quantum}\ }\textbf {\bibinfo {volume} {2}},\ \bibinfo {pages}
  {020310} (\bibinfo {year} {2021})}\BibitemShut {NoStop}%
\bibitem [{\citenamefont {Sureshbabu}\ \emph {et~al.}(2021)\citenamefont
  {Sureshbabu}, \citenamefont {Sajjan}, \citenamefont {Oh},\ and\ \citenamefont
  {Kais}}]{sureshbabu2021}%
  \BibitemOpen
  \bibfield  {author} {\bibinfo {author} {\bibfnamefont {S.~H.}\ \bibnamefont
  {Sureshbabu}}, \bibinfo {author} {\bibfnamefont {M.}~\bibnamefont {Sajjan}},
  \bibinfo {author} {\bibfnamefont {S.}~\bibnamefont {Oh}},\ and\ \bibinfo
  {author} {\bibfnamefont {S.}~\bibnamefont {Kais}},\ }\bibfield  {title}
  {\bibinfo {title} {Implementation of quantum machine learning for electronic
  structure calculations of periodic systems on quantum computing devices},\
  }\href {https://doi.org/10.1021/acs.jcim.1c00294} {\bibfield  {journal}
  {\bibinfo  {journal} {J. Chem. Inf. Model.}\ }\textbf {\bibinfo {volume}
  {61}},\ \bibinfo {pages} {2667} (\bibinfo {year} {2021})}\BibitemShut
  {NoStop}%
\bibitem [{\citenamefont {Smart}\ \emph
  {et~al.}(2022{\natexlab{a}})\citenamefont {Smart}, \citenamefont {Boyn},\
  and\ \citenamefont {Mazziotti}}]{smart2022benzyne}%
  \BibitemOpen
  \bibfield  {author} {\bibinfo {author} {\bibfnamefont {S.~E.}\ \bibnamefont
  {Smart}}, \bibinfo {author} {\bibfnamefont {J.-N.}\ \bibnamefont {Boyn}},\
  and\ \bibinfo {author} {\bibfnamefont {D.~A.}\ \bibnamefont {Mazziotti}},\
  }\bibfield  {title} {\bibinfo {title} {Resolving correlated states of benzyne
  with an error-mitigated contracted quantum eigensolver},\ }\href
  {https://doi.org/10.1103/PhysRevA.105.022405} {\bibfield  {journal} {\bibinfo
   {journal} {Phys. Rev. A}\ }\textbf {\bibinfo {volume} {105}},\ \bibinfo
  {pages} {022405} (\bibinfo {year} {2022}{\natexlab{a}})}\BibitemShut
  {NoStop}%
\bibitem [{\citenamefont {Dutta}\ \emph
  {et~al.}(2024{\natexlab{b}})\citenamefont {Dutta}, \citenamefont {Vu},
  \citenamefont {Lyu}, \citenamefont {Wang},\ and\ \citenamefont
  {Batista}}]{dutta20242}%
  \BibitemOpen
  \bibfield  {author} {\bibinfo {author} {\bibfnamefont {R.}~\bibnamefont
  {Dutta}}, \bibinfo {author} {\bibfnamefont {N.~P.}\ \bibnamefont {Vu}},
  \bibinfo {author} {\bibfnamefont {N.}~\bibnamefont {Lyu}}, \bibinfo {author}
  {\bibfnamefont {C.}~\bibnamefont {Wang}},\ and\ \bibinfo {author}
  {\bibfnamefont {V.~S.}\ \bibnamefont {Batista}},\ }\bibfield  {title}
  {\bibinfo {title} {Simulating electronic structure on bosonic quantum
  computers},\ }\href@noop {} {\bibfield  {journal} {\bibinfo  {journal} {arXiv
  preprint arXiv:2404.10222}\ } (\bibinfo {year}
  {2024}{\natexlab{b}})}\BibitemShut {NoStop}%
\bibitem [{\citenamefont {Cianci}\ \emph {et~al.}(2024)\citenamefont {Cianci},
  \citenamefont {Santos},\ and\ \citenamefont
  {Batista}}]{cianci2024subspacesearchquantumimaginarytime}%
  \BibitemOpen
  \bibfield  {author} {\bibinfo {author} {\bibfnamefont {C.}~\bibnamefont
  {Cianci}}, \bibinfo {author} {\bibfnamefont {L.~F.}\ \bibnamefont {Santos}},\
  and\ \bibinfo {author} {\bibfnamefont {V.~S.}\ \bibnamefont {Batista}},\
  }\href {https://arxiv.org/abs/2407.11182} {\bibinfo {title} {Subspace-search
  quantum imaginary time evolution for excited state computations}} (\bibinfo
  {year} {2024}),\ \Eprint {https://arxiv.org/abs/2407.11182} {arXiv:2407.11182
  [quant-ph]} \BibitemShut {NoStop}%
\bibitem [{\citenamefont {Cao}\ \emph {et~al.}(2018)\citenamefont {Cao},
  \citenamefont {Romero},\ and\ \citenamefont {Aspuru-Guzik}}]{Cao2018}%
  \BibitemOpen
  \bibfield  {author} {\bibinfo {author} {\bibfnamefont {Y.}~\bibnamefont
  {Cao}}, \bibinfo {author} {\bibfnamefont {J.}~\bibnamefont {Romero}},\ and\
  \bibinfo {author} {\bibfnamefont {A.}~\bibnamefont {Aspuru-Guzik}},\
  }\bibfield  {title} {\bibinfo {title} {Potential of quantum computing for
  drug discovery},\ }\href {https://doi.org/10.1147/JRD.2018.2888987}
  {\bibfield  {journal} {\bibinfo  {journal} {IBM Journal of Research and
  Development}\ }\textbf {\bibinfo {volume} {62}},\ \bibinfo {pages} {6:1}
  (\bibinfo {year} {2018})}\BibitemShut {NoStop}%
\bibitem [{\citenamefont {Smaldone}\ and\ \citenamefont
  {Batista}()}]{smaldone2024}%
  \BibitemOpen
  \bibfield  {author} {\bibinfo {author} {\bibfnamefont {A.~M.}\ \bibnamefont
  {Smaldone}}\ and\ \bibinfo {author} {\bibfnamefont {V.~S.}\ \bibnamefont
  {Batista}},\ }\bibfield  {title} {\bibinfo {title} {Quantum-to-classical
  neural network transfer learning applied to drug toxicity prediction},\
  }\href {https://doi.org/10.1021/acs.jctc.4c00432} {\bibinfo  {journal} {J.
  Chem. Theory Comput.}\ ,\ \bibinfo {pages} {4901}}\BibitemShut {NoStop}%
\bibitem [{\citenamefont {Varsano}\ \emph {et~al.}(2006)\citenamefont
  {Varsano}, \citenamefont {Di~Felice}, \citenamefont {Marques},\ and\
  \citenamefont {Rubio}}]{varsano2006}%
  \BibitemOpen
\bibfield  {journal} {  }\bibfield  {author} {\bibinfo {author} {\bibfnamefont
  {D.}~\bibnamefont {Varsano}}, \bibinfo {author} {\bibfnamefont
  {R.}~\bibnamefont {Di~Felice}}, \bibinfo {author} {\bibfnamefont {M.~A.}\
  \bibnamefont {Marques}},\ and\ \bibinfo {author} {\bibfnamefont
  {A.}~\bibnamefont {Rubio}},\ }\bibfield  {title} {\bibinfo {title} {A {TDDFT}
  study of the excited states of {DNA} bases and their assemblies},\ }\href
  {https://doi.org/10.1021/jp056120g} {\bibfield  {journal} {\bibinfo
  {journal} {J. Phys. Chem. B}\ }\textbf {\bibinfo {volume} {110}},\ \bibinfo
  {pages} {7129} (\bibinfo {year} {2006})}\BibitemShut {NoStop}%
\bibitem [{\citenamefont {Teh}\ and\ \citenamefont {Subotnik}(2019)}]{teh2019}%
  \BibitemOpen
  \bibfield  {author} {\bibinfo {author} {\bibfnamefont {H.-H.}\ \bibnamefont
  {Teh}}\ and\ \bibinfo {author} {\bibfnamefont {J.~E.}\ \bibnamefont
  {Subotnik}},\ }\bibfield  {title} {\bibinfo {title} {The simplest possible
  approach for simulating {S$_0$}--{S$_1$} conical intersections with
  {DFT/TDDFT}: Adding one doubly excited configuration},\ }\href
  {https://doi.org/10.1021/acs.jpclett.9b00981} {\bibfield  {journal} {\bibinfo
   {journal} {J. Phys. Chem. Lett.}\ }\textbf {\bibinfo {volume} {10}},\
  \bibinfo {pages} {3426} (\bibinfo {year} {2019})}\BibitemShut {NoStop}%
\bibitem [{\citenamefont {Marrot}\ and\ \citenamefont
  {Meunier}(2008)}]{marrot2008}%
  \BibitemOpen
  \bibfield  {author} {\bibinfo {author} {\bibfnamefont {L.}~\bibnamefont
  {Marrot}}\ and\ \bibinfo {author} {\bibfnamefont {J.-R.}\ \bibnamefont
  {Meunier}},\ }\bibfield  {title} {\bibinfo {title} {Skin dna photodamage and
  its biological consequences},\ }\href@noop {} {\bibfield  {journal} {\bibinfo
   {journal} {J. Am. Acad. Dermatol.}\ }\textbf {\bibinfo {volume} {58}},\
  \bibinfo {pages} {S139} (\bibinfo {year} {2008})}\BibitemShut {NoStop}%
\bibitem [{\citenamefont {Kang}\ \emph {et~al.}(2002)\citenamefont {Kang},
  \citenamefont {Lee}, \citenamefont {Jung}, \citenamefont {Ko},\ and\
  \citenamefont {Kim}}]{kang2002}%
  \BibitemOpen
  \bibfield  {author} {\bibinfo {author} {\bibfnamefont {H.}~\bibnamefont
  {Kang}}, \bibinfo {author} {\bibfnamefont {K.~T.}\ \bibnamefont {Lee}},
  \bibinfo {author} {\bibfnamefont {B.}~\bibnamefont {Jung}}, \bibinfo {author}
  {\bibfnamefont {Y.~J.}\ \bibnamefont {Ko}},\ and\ \bibinfo {author}
  {\bibfnamefont {S.~K.}\ \bibnamefont {Kim}},\ }\bibfield  {title} {\bibinfo
  {title} {Intrinsic lifetimes of the excited state of {DNA} and {RNA} bases},\
  }\href {https://doi.org/10.1021/ja027627x} {\bibfield  {journal} {\bibinfo
  {journal} {J. Am. Chem. Soc.}\ }\textbf {\bibinfo {volume} {124}},\ \bibinfo
  {pages} {12958} (\bibinfo {year} {2002})}\BibitemShut {NoStop}%
\bibitem [{\citenamefont {Malone}\ \emph {et~al.}(2003)\citenamefont {Malone},
  \citenamefont {Miller},\ and\ \citenamefont {Kohler}}]{malone2003}%
  \BibitemOpen
  \bibfield  {author} {\bibinfo {author} {\bibfnamefont {R.~J.}\ \bibnamefont
  {Malone}}, \bibinfo {author} {\bibfnamefont {A.~M.}\ \bibnamefont {Miller}},\
  and\ \bibinfo {author} {\bibfnamefont {B.}~\bibnamefont {Kohler}},\
  }\bibfield  {title} {\bibinfo {title} {Singlet excited-state lifetimes of
  cytosine derivatives measured by femtosecond transient absorption},\ }\href
  {https://doi.org/10.1562/0031-8655(2003)0770158SESLOC2.0.CO2} {\bibfield
  {journal} {\bibinfo  {journal} {Photochem. Photobiol.}\ }\textbf {\bibinfo
  {volume} {77}},\ \bibinfo {pages} {158} (\bibinfo {year} {2003})}\BibitemShut
  {NoStop}%
\bibitem [{\citenamefont {Sharonov}\ \emph {et~al.}(2003)\citenamefont
  {Sharonov}, \citenamefont {Gustavsson}, \citenamefont {Carr{\'e}},
  \citenamefont {Renault},\ and\ \citenamefont {Markovitsi}}]{sharonov2003}%
  \BibitemOpen
  \bibfield  {author} {\bibinfo {author} {\bibfnamefont {A.}~\bibnamefont
  {Sharonov}}, \bibinfo {author} {\bibfnamefont {T.}~\bibnamefont
  {Gustavsson}}, \bibinfo {author} {\bibfnamefont {V.}~\bibnamefont
  {Carr{\'e}}}, \bibinfo {author} {\bibfnamefont {E.}~\bibnamefont {Renault}},\
  and\ \bibinfo {author} {\bibfnamefont {D.}~\bibnamefont {Markovitsi}},\
  }\bibfield  {title} {\bibinfo {title} {Cytosine excited state dynamics
  studied by femtosecond fluorescence upconversion and transient absorption
  spectroscopy},\ }\href {https://doi.org/10.1016/j.cplett.2003.09.021}
  {\bibfield  {journal} {\bibinfo  {journal} {Chem. Phys. Lett.}\ }\textbf
  {\bibinfo {volume} {380}},\ \bibinfo {pages} {173} (\bibinfo {year}
  {2003})}\BibitemShut {NoStop}%
\bibitem [{\citenamefont {Merch{\'a}n}\ and\ \citenamefont
  {Serrano-Andr{\'e}s}(2003)}]{merchan2003}%
  \BibitemOpen
  \bibfield  {author} {\bibinfo {author} {\bibfnamefont {M.}~\bibnamefont
  {Merch{\'a}n}}\ and\ \bibinfo {author} {\bibfnamefont {L.}~\bibnamefont
  {Serrano-Andr{\'e}s}},\ }\bibfield  {title} {\bibinfo {title} {Ultrafast
  internal conversion of excited cytosine via the lowest $\pi$$\pi$* electronic
  singlet state},\ }\href {https://doi.org/10.1021/ja0351600} {\bibfield
  {journal} {\bibinfo  {journal} {J. Am. Chem. Soc.}\ }\textbf {\bibinfo
  {volume} {125}},\ \bibinfo {pages} {8108} (\bibinfo {year}
  {2003})}\BibitemShut {NoStop}%
\bibitem [{\citenamefont {Blancafort}\ and\ \citenamefont
  {Robb}(2004)}]{blancafort2004}%
  \BibitemOpen
  \bibfield  {author} {\bibinfo {author} {\bibfnamefont {L.}~\bibnamefont
  {Blancafort}}\ and\ \bibinfo {author} {\bibfnamefont {M.~A.}\ \bibnamefont
  {Robb}},\ }\bibfield  {title} {\bibinfo {title} {Key role of a threefold
  state crossing in the ultrafast decay of electronically excited cytosine},\
  }\href {https://doi.org/10.1021/jp045985b} {\bibfield  {journal} {\bibinfo
  {journal} {J. Phys. Chem. A}\ }\textbf {\bibinfo {volume} {108}},\ \bibinfo
  {pages} {10609} (\bibinfo {year} {2004})}\BibitemShut {NoStop}%
\bibitem [{\citenamefont {Matsika}(2005)}]{matsika2005}%
  \BibitemOpen
  \bibfield  {author} {\bibinfo {author} {\bibfnamefont {S.}~\bibnamefont
  {Matsika}},\ }\bibfield  {title} {\bibinfo {title} {Three-state conical
  intersections in nucleic acid bases},\ }\href
  {https://doi.org/10.1021/jp0513622} {\bibfield  {journal} {\bibinfo
  {journal} {J. Phys. Chem. A}\ }\textbf {\bibinfo {volume} {109}},\ \bibinfo
  {pages} {7538} (\bibinfo {year} {2005})}\BibitemShut {NoStop}%
\bibitem [{\citenamefont {Kistler}\ and\ \citenamefont
  {Matsika}(2008)}]{kistler2008}%
  \BibitemOpen
  \bibfield  {author} {\bibinfo {author} {\bibfnamefont {K.~A.}\ \bibnamefont
  {Kistler}}\ and\ \bibinfo {author} {\bibfnamefont {S.}~\bibnamefont
  {Matsika}},\ }\bibfield  {title} {\bibinfo {title} {Three-state conical
  intersections in cytosine and pyrimidinone bases},\ }\href
  {https://doi.org/10.1063/1.2932102} {\bibfield  {journal} {\bibinfo
  {journal} {J. Chem. Phys.}\ }\textbf {\bibinfo {volume} {128}},\ \bibinfo
  {pages} {215102} (\bibinfo {year} {2008})}\BibitemShut {NoStop}%
\bibitem [{\citenamefont {Gonz{\'a}lez-V{\'a}zquez}\ and\ \citenamefont
  {Gonz{\'a}lez}(2010)}]{gonzalez2010}%
  \BibitemOpen
  \bibfield  {author} {\bibinfo {author} {\bibfnamefont {J.}~\bibnamefont
  {Gonz{\'a}lez-V{\'a}zquez}}\ and\ \bibinfo {author} {\bibfnamefont
  {L.}~\bibnamefont {Gonz{\'a}lez}},\ }\bibfield  {title} {\bibinfo {title} {A
  time-dependent picture of the ultrafast deactivation of keto-cytosine
  including three-state conical intersections},\ }\href
  {https://doi.org/10.1002/cphc.201000557} {\bibfield  {journal} {\bibinfo
  {journal} {ChemPhysChem}\ }\textbf {\bibinfo {volume} {11}},\ \bibinfo
  {pages} {3617} (\bibinfo {year} {2010})}\BibitemShut {NoStop}%
\bibitem [{\citenamefont {Barbatti}\ \emph {et~al.}(2011)\citenamefont
  {Barbatti}, \citenamefont {Aquino}, \citenamefont {Szymczak}, \citenamefont
  {Nachtigallova},\ and\ \citenamefont {Lischka}}]{barbatti2011}%
  \BibitemOpen
  \bibfield  {author} {\bibinfo {author} {\bibfnamefont {M.}~\bibnamefont
  {Barbatti}}, \bibinfo {author} {\bibfnamefont {A.~J.}\ \bibnamefont
  {Aquino}}, \bibinfo {author} {\bibfnamefont {J.~J.}\ \bibnamefont
  {Szymczak}}, \bibinfo {author} {\bibfnamefont {D.}~\bibnamefont
  {Nachtigallova}},\ and\ \bibinfo {author} {\bibfnamefont {H.}~\bibnamefont
  {Lischka}},\ }\bibfield  {title} {\bibinfo {title} {Photodynamical
  simulations of cytosine: characterization of the ultrafast bi-exponential uv
  deactivation},\ }\href {https://doi.org/10.1039/C0CP01327G} {\bibfield
  {journal} {\bibinfo  {journal} {Phys. Chem. Chem. Phys.}\ }\textbf {\bibinfo
  {volume} {13}},\ \bibinfo {pages} {6145} (\bibinfo {year}
  {2011})}\BibitemShut {NoStop}%
\bibitem [{\citenamefont {Richter}\ \emph {et~al.}(2012)\citenamefont
  {Richter}, \citenamefont {Marquetand}, \citenamefont {Gonzalez-Vazquez},
  \citenamefont {Sola},\ and\ \citenamefont {Gonz{\'a}lez}}]{richter2012}%
  \BibitemOpen
  \bibfield  {author} {\bibinfo {author} {\bibfnamefont {M.}~\bibnamefont
  {Richter}}, \bibinfo {author} {\bibfnamefont {P.}~\bibnamefont {Marquetand}},
  \bibinfo {author} {\bibfnamefont {J.}~\bibnamefont {Gonzalez-Vazquez}},
  \bibinfo {author} {\bibfnamefont {I.}~\bibnamefont {Sola}},\ and\ \bibinfo
  {author} {\bibfnamefont {L.}~\bibnamefont {Gonz{\'a}lez}},\ }\bibfield
  {title} {\bibinfo {title} {Femtosecond intersystem crossing in the {DNA}
  nucleobase cytosine},\ }\href {https://doi.org/10.1021/jz301312h} {\bibfield
  {journal} {\bibinfo  {journal} {J. Phys. Chem. Lett}\ }\textbf {\bibinfo
  {volume} {3}},\ \bibinfo {pages} {3090} (\bibinfo {year} {2012})}\BibitemShut
  {NoStop}%
\bibitem [{\citenamefont {Blaser}\ \emph {et~al.}(2016)\citenamefont {Blaser},
  \citenamefont {Trachsel}, \citenamefont {Lobsiger}, \citenamefont {Wiedmer},
  \citenamefont {Frey},\ and\ \citenamefont {Leutwyler}}]{blaser2016}%
  \BibitemOpen
  \bibfield  {author} {\bibinfo {author} {\bibfnamefont {S.}~\bibnamefont
  {Blaser}}, \bibinfo {author} {\bibfnamefont {M.~A.}\ \bibnamefont
  {Trachsel}}, \bibinfo {author} {\bibfnamefont {S.}~\bibnamefont {Lobsiger}},
  \bibinfo {author} {\bibfnamefont {T.}~\bibnamefont {Wiedmer}}, \bibinfo
  {author} {\bibfnamefont {H.-M.}\ \bibnamefont {Frey}},\ and\ \bibinfo
  {author} {\bibfnamefont {S.}~\bibnamefont {Leutwyler}},\ }\bibfield  {title}
  {\bibinfo {title} {Gas-phase cytosine and cytosine-n1-derivatives have 0.1--1
  ns lifetimes near the s1 state minimum},\ }\href
  {https://doi.org/10.1021/acs.jpclett.6b00017} {\bibfield  {journal} {\bibinfo
   {journal} {J. Phys. Chem. Lett.}\ }\textbf {\bibinfo {volume} {7}},\
  \bibinfo {pages} {752} (\bibinfo {year} {2016})}\BibitemShut {NoStop}%
\bibitem [{\citenamefont {Trachsel}\ \emph {et~al.}(2020)\citenamefont
  {Trachsel}, \citenamefont {Blaser}, \citenamefont {Lobsiger}, \citenamefont
  {Siffert}, \citenamefont {Frey}, \citenamefont {Blancafort},\ and\
  \citenamefont {Leutwyler}}]{trachsel2020}%
  \BibitemOpen
  \bibfield  {author} {\bibinfo {author} {\bibfnamefont {M.~A.}\ \bibnamefont
  {Trachsel}}, \bibinfo {author} {\bibfnamefont {S.}~\bibnamefont {Blaser}},
  \bibinfo {author} {\bibfnamefont {S.}~\bibnamefont {Lobsiger}}, \bibinfo
  {author} {\bibfnamefont {L.}~\bibnamefont {Siffert}}, \bibinfo {author}
  {\bibfnamefont {H.-M.}\ \bibnamefont {Frey}}, \bibinfo {author}
  {\bibfnamefont {L.}~\bibnamefont {Blancafort}},\ and\ \bibinfo {author}
  {\bibfnamefont {S.}~\bibnamefont {Leutwyler}},\ }\bibfield  {title} {\bibinfo
  {title} {Locating cytosine conical intersections by laser experiments and ab
  initio calculations},\ }\href {https://doi.org/10.1021/acs.jpclett.0c00779}
  {\bibfield  {journal} {\bibinfo  {journal} {J. Phys. Chem. Lett.}\ }\textbf
  {\bibinfo {volume} {11}},\ \bibinfo {pages} {3203} (\bibinfo {year}
  {2020})}\BibitemShut {NoStop}%
\bibitem [{\citenamefont {Shahrokh}\ \emph {et~al.}(2021)\citenamefont
  {Shahrokh}, \citenamefont {Omidyan},\ and\ \citenamefont
  {Azimi}}]{shahrokh2021}%
  \BibitemOpen
  \bibfield  {author} {\bibinfo {author} {\bibfnamefont {L.}~\bibnamefont
  {Shahrokh}}, \bibinfo {author} {\bibfnamefont {R.}~\bibnamefont {Omidyan}},\
  and\ \bibinfo {author} {\bibfnamefont {G.}~\bibnamefont {Azimi}},\ }\bibfield
   {title} {\bibinfo {title} {Theoretical insights on the
  excited-state-deactivation mechanisms of protonated thymine and cytosine},\
  }\href {https://doi.org/10.1039/D0CP06673G} {\bibfield  {journal} {\bibinfo
  {journal} {Phys. Chem. Chem. Phys.}\ }\textbf {\bibinfo {volume} {23}},\
  \bibinfo {pages} {8916} (\bibinfo {year} {2021})}\BibitemShut {NoStop}%
\bibitem [{\citenamefont {Yarkony}(1996)}]{yarkony1996}%
  \BibitemOpen
  \bibfield  {author} {\bibinfo {author} {\bibfnamefont {D.~R.}\ \bibnamefont
  {Yarkony}},\ }\bibfield  {title} {\bibinfo {title} {Diabolical conical
  intersections},\ }\href {https://doi.org/10.1103/RevModPhys.68.985}
  {\bibfield  {journal} {\bibinfo  {journal} {Rev. Mod. Phys.}\ }\textbf
  {\bibinfo {volume} {68}},\ \bibinfo {pages} {985} (\bibinfo {year}
  {1996})}\BibitemShut {NoStop}%
\bibitem [{\citenamefont {Domcke}\ \emph {et~al.}(2004)\citenamefont {Domcke},
  \citenamefont {Yarkony},\ and\ \citenamefont {K{\"o}ppel}}]{domcke2004}%
  \BibitemOpen
  \bibfield  {author} {\bibinfo {author} {\bibfnamefont {W.}~\bibnamefont
  {Domcke}}, \bibinfo {author} {\bibfnamefont {D.}~\bibnamefont {Yarkony}},\
  and\ \bibinfo {author} {\bibfnamefont {H.}~\bibnamefont {K{\"o}ppel}},\
  }\href@noop {} {\emph {\bibinfo {title} {Conical intersections: Electronic
  structure, dynamics \& spectroscopy}}},\ Vol.~\bibinfo {volume} {15}\
  (\bibinfo  {publisher} {World Scientific},\ \bibinfo {year}
  {2004})\BibitemShut {NoStop}%
\bibitem [{\citenamefont {Levine}\ and\ \citenamefont
  {Mart{\'\i}nez}(2007)}]{levine2007}%
  \BibitemOpen
  \bibfield  {author} {\bibinfo {author} {\bibfnamefont {B.~G.}\ \bibnamefont
  {Levine}}\ and\ \bibinfo {author} {\bibfnamefont {T.~J.}\ \bibnamefont
  {Mart{\'\i}nez}},\ }\bibfield  {title} {\bibinfo {title} {Isomerization
  through conical intersections},\ }\href
  {https://doi.org/10.1146/annurev.physchem.57.032905.104612} {\bibfield
  {journal} {\bibinfo  {journal} {Annu. Rev. Phys. Chem.}\ }\textbf {\bibinfo
  {volume} {58}},\ \bibinfo {pages} {613} (\bibinfo {year} {2007})}\BibitemShut
  {NoStop}%
\bibitem [{\citenamefont {Matsika}\ and\ \citenamefont
  {Krause}(2011)}]{matsika2011}%
  \BibitemOpen
  \bibfield  {author} {\bibinfo {author} {\bibfnamefont {S.}~\bibnamefont
  {Matsika}}\ and\ \bibinfo {author} {\bibfnamefont {P.}~\bibnamefont
  {Krause}},\ }\bibfield  {title} {\bibinfo {title} {Nonadiabatic events and
  conical intersections},\ }\href
  {https://doi.org/10.1146/annurev-physchem-032210-103450} {\bibfield
  {journal} {\bibinfo  {journal} {Annu. Rev. Phys. Chem.}\ }\textbf {\bibinfo
  {volume} {62}},\ \bibinfo {pages} {621} (\bibinfo {year} {2011})}\BibitemShut
  {NoStop}%
\bibitem [{\citenamefont {Tully}(2012)}]{tully2012}%
  \BibitemOpen
  \bibfield  {author} {\bibinfo {author} {\bibfnamefont {J.~C.}\ \bibnamefont
  {Tully}},\ }\bibfield  {title} {\bibinfo {title} {Perspective: Nonadiabatic
  dynamics theory},\ }\href {https://doi.org/10.1063/1.4757762} {\bibfield
  {journal} {\bibinfo  {journal} {J. Chem. Phys.}\ }\textbf {\bibinfo {volume}
  {137}} (\bibinfo {year} {2012})}\BibitemShut {NoStop}%
\bibitem [{\citenamefont {Guo}\ and\ \citenamefont {Yarkony}(2016)}]{guo2016}%
  \BibitemOpen
  \bibfield  {author} {\bibinfo {author} {\bibfnamefont {H.}~\bibnamefont
  {Guo}}\ and\ \bibinfo {author} {\bibfnamefont {D.~R.}\ \bibnamefont
  {Yarkony}},\ }\bibfield  {title} {\bibinfo {title} {Accurate nonadiabatic
  dynamics},\ }\href {https://doi.org/10.1039/C6CP05553B} {\bibfield  {journal}
  {\bibinfo  {journal} {Phys. Chem. Chem. Phys.}\ }\textbf {\bibinfo {volume}
  {18}},\ \bibinfo {pages} {26335} (\bibinfo {year} {2016})}\BibitemShut
  {NoStop}%
\bibitem [{\citenamefont {Yarkony}\ \emph {et~al.}(2019)\citenamefont
  {Yarkony}, \citenamefont {Xie}, \citenamefont {Zhu}, \citenamefont {Wang},
  \citenamefont {Malbon},\ and\ \citenamefont {Guo}}]{yarkony2019}%
  \BibitemOpen
  \bibfield  {author} {\bibinfo {author} {\bibfnamefont {D.~R.}\ \bibnamefont
  {Yarkony}}, \bibinfo {author} {\bibfnamefont {C.}~\bibnamefont {Xie}},
  \bibinfo {author} {\bibfnamefont {X.}~\bibnamefont {Zhu}}, \bibinfo {author}
  {\bibfnamefont {Y.}~\bibnamefont {Wang}}, \bibinfo {author} {\bibfnamefont
  {C.~L.}\ \bibnamefont {Malbon}},\ and\ \bibinfo {author} {\bibfnamefont
  {H.}~\bibnamefont {Guo}},\ }\bibfield  {title} {\bibinfo {title} {Diabatic
  and adiabatic representations: Electronic structure caveats},\ }\href
  {https://doi.org/10.1016/j.comptc.2019.01.020} {\bibfield  {journal}
  {\bibinfo  {journal} {Comput. Theor. Chem.}\ }\textbf {\bibinfo {volume}
  {1152}},\ \bibinfo {pages} {41} (\bibinfo {year} {2019})}\BibitemShut
  {NoStop}%
\bibitem [{\citenamefont {Yalouz}\ \emph {et~al.}(2021)\citenamefont {Yalouz},
  \citenamefont {Senjean}, \citenamefont {G{\"u}nther}, \citenamefont {Buda},
  \citenamefont {O’Brien},\ and\ \citenamefont {Visscher}}]{yalouz2021}%
  \BibitemOpen
  \bibfield  {author} {\bibinfo {author} {\bibfnamefont {S.}~\bibnamefont
  {Yalouz}}, \bibinfo {author} {\bibfnamefont {B.}~\bibnamefont {Senjean}},
  \bibinfo {author} {\bibfnamefont {J.}~\bibnamefont {G{\"u}nther}}, \bibinfo
  {author} {\bibfnamefont {F.}~\bibnamefont {Buda}}, \bibinfo {author}
  {\bibfnamefont {T.~E.}\ \bibnamefont {O’Brien}},\ and\ \bibinfo {author}
  {\bibfnamefont {L.}~\bibnamefont {Visscher}},\ }\bibfield  {title} {\bibinfo
  {title} {A state-averaged orbital-optimized hybrid quantum--classical
  algorithm for a democratic description of ground and excited states},\ }\href
  {https://doi.org/10.1088/2058-9565/abd334} {\bibfield  {journal} {\bibinfo
  {journal} {Quantum Sci. Technol.}\ }\textbf {\bibinfo {volume} {6}},\
  \bibinfo {pages} {024004} (\bibinfo {year} {2021})}\BibitemShut {NoStop}%
\bibitem [{\citenamefont {Ollitrault}\ \emph {et~al.}(2020)\citenamefont
  {Ollitrault}, \citenamefont {Mazzola},\ and\ \citenamefont
  {Tavernelli}}]{ollitrault2020}%
  \BibitemOpen
  \bibfield  {author} {\bibinfo {author} {\bibfnamefont {P.~J.}\ \bibnamefont
  {Ollitrault}}, \bibinfo {author} {\bibfnamefont {G.}~\bibnamefont
  {Mazzola}},\ and\ \bibinfo {author} {\bibfnamefont {I.}~\bibnamefont
  {Tavernelli}},\ }\bibfield  {title} {\bibinfo {title} {Nonadiabatic molecular
  quantum dynamics with quantum computers},\ }\href
  {https://doi.org/10.1103/PhysRevLett.125.260511} {\bibfield  {journal}
  {\bibinfo  {journal} {Phys. Rev. Lett.}\ }\textbf {\bibinfo {volume} {125}},\
  \bibinfo {pages} {260511} (\bibinfo {year} {2020})}\BibitemShut {NoStop}%
\bibitem [{\citenamefont {Wang}\ and\ \citenamefont
  {Mazziotti}(2024)}]{wang2024}%
  \BibitemOpen
  \bibfield  {author} {\bibinfo {author} {\bibfnamefont {Y.}~\bibnamefont
  {Wang}}\ and\ \bibinfo {author} {\bibfnamefont {D.~A.}\ \bibnamefont
  {Mazziotti}},\ }\bibfield  {title} {\bibinfo {title} {Quantum simulation of
  conical intersections},\ }\href {https://doi.org/10.1039/D4CP00391H}
  {\bibfield  {journal} {\bibinfo  {journal} {Phys. Chem. Chem. Phys.}\
  }\textbf {\bibinfo {volume} {26}},\ \bibinfo {pages} {11491} (\bibinfo {year}
  {2024})}\BibitemShut {NoStop}%
\bibitem [{\citenamefont {Koridon}\ \emph {et~al.}(2024)\citenamefont
  {Koridon}, \citenamefont {Fraxanet}, \citenamefont {Dauphin}, \citenamefont
  {Visscher}, \citenamefont {O'Brien},\ and\ \citenamefont
  {Polla}}]{koridon2024}%
  \BibitemOpen
  \bibfield  {author} {\bibinfo {author} {\bibfnamefont {E.}~\bibnamefont
  {Koridon}}, \bibinfo {author} {\bibfnamefont {J.}~\bibnamefont {Fraxanet}},
  \bibinfo {author} {\bibfnamefont {A.}~\bibnamefont {Dauphin}}, \bibinfo
  {author} {\bibfnamefont {L.}~\bibnamefont {Visscher}}, \bibinfo {author}
  {\bibfnamefont {T.~E.}\ \bibnamefont {O'Brien}},\ and\ \bibinfo {author}
  {\bibfnamefont {S.}~\bibnamefont {Polla}},\ }\bibfield  {title} {\bibinfo
  {title} {A hybrid quantum algorithm to detect conical intersections},\ }\href
  {https://doi.org/10.22331/q-2024-02-20-1259} {\bibfield  {journal} {\bibinfo
  {journal} {Quantum}\ }\textbf {\bibinfo {volume} {8}},\ \bibinfo {pages}
  {1259} (\bibinfo {year} {2024})}\BibitemShut {NoStop}%
\bibitem [{\citenamefont {Zhao}\ \emph {et~al.}()\citenamefont {Zhao},
  \citenamefont {Tang}, \citenamefont {Xiao}, \citenamefont {Wang},
  \citenamefont {Sun}, \citenamefont {Chen}, \citenamefont {Cai}, \citenamefont
  {Li}, \citenamefont {Yu},\ and\ \citenamefont {Fang}}]{zhao2024}%
  \BibitemOpen
  \bibfield  {author} {\bibinfo {author} {\bibfnamefont {S.}~\bibnamefont
  {Zhao}}, \bibinfo {author} {\bibfnamefont {D.}~\bibnamefont {Tang}}, \bibinfo
  {author} {\bibfnamefont {X.}~\bibnamefont {Xiao}}, \bibinfo {author}
  {\bibfnamefont {R.}~\bibnamefont {Wang}}, \bibinfo {author} {\bibfnamefont
  {Q.}~\bibnamefont {Sun}}, \bibinfo {author} {\bibfnamefont {Z.}~\bibnamefont
  {Chen}}, \bibinfo {author} {\bibfnamefont {X.}~\bibnamefont {Cai}}, \bibinfo
  {author} {\bibfnamefont {Z.}~\bibnamefont {Li}}, \bibinfo {author}
  {\bibfnamefont {H.}~\bibnamefont {Yu}},\ and\ \bibinfo {author}
  {\bibfnamefont {W.-H.}\ \bibnamefont {Fang}},\ }\bibfield  {title} {\bibinfo
  {title} {Quantum computation of conical intersections on a programmable
  superconducting quantum processor},\ }\href
  {https://doi.org/10.1021/acs.jpclett.4c01314} {\bibfield  {journal} {\bibinfo
   {journal} {J. Phys. Chem. Lett.}\ }\textbf {\bibinfo {volume} {15}},\
  \bibinfo {pages} {7244}}\BibitemShut {NoStop}%
\bibitem [{\citenamefont {Higgott}\ \emph {et~al.}(2019)\citenamefont
  {Higgott}, \citenamefont {Wang},\ and\ \citenamefont
  {Brierley}}]{higgott2019}%
  \BibitemOpen
  \bibfield  {author} {\bibinfo {author} {\bibfnamefont {O.}~\bibnamefont
  {Higgott}}, \bibinfo {author} {\bibfnamefont {D.}~\bibnamefont {Wang}},\ and\
  \bibinfo {author} {\bibfnamefont {S.}~\bibnamefont {Brierley}},\ }\bibfield
  {title} {\bibinfo {title} {Variational quantum computation of excited
  states},\ }\href {https://doi.org/10.22331/q-2019-07-01-156} {\bibfield
  {journal} {\bibinfo  {journal} {Quantum}\ }\textbf {\bibinfo {volume} {3}},\
  \bibinfo {pages} {156} (\bibinfo {year} {2019})}\BibitemShut {NoStop}%
\bibitem [{\citenamefont {Smart}\ and\ \citenamefont
  {Mazziotti}(2021)}]{smart2021}%
  \BibitemOpen
  \bibfield  {author} {\bibinfo {author} {\bibfnamefont {S.~E.}\ \bibnamefont
  {Smart}}\ and\ \bibinfo {author} {\bibfnamefont {D.~A.}\ \bibnamefont
  {Mazziotti}},\ }\bibfield  {title} {\bibinfo {title} {Quantum solver of
  contracted eigenvalue equations for scalable molecular simulations on quantum
  computing devices},\ }\href {https://doi.org/10.1103/PhysRevLett.126.070504}
  {\bibfield  {journal} {\bibinfo  {journal} {Phys. Rev. Lett.}\ }\textbf
  {\bibinfo {volume} {126}},\ \bibinfo {pages} {070504} (\bibinfo {year}
  {2021})}\BibitemShut {NoStop}%
\bibitem [{\citenamefont {Smart}\ \emph
  {et~al.}(2022{\natexlab{b}})\citenamefont {Smart}, \citenamefont {Boyn},\
  and\ \citenamefont {Mazziotti}}]{smart2022}%
  \BibitemOpen
  \bibfield  {author} {\bibinfo {author} {\bibfnamefont {S.~E.}\ \bibnamefont
  {Smart}}, \bibinfo {author} {\bibfnamefont {J.-N.}\ \bibnamefont {Boyn}},\
  and\ \bibinfo {author} {\bibfnamefont {D.~A.}\ \bibnamefont {Mazziotti}},\
  }\bibfield  {title} {\bibinfo {title} {Resolving correlated states of benzyne
  with an error-mitigated contracted quantum eigensolver},\ }\href
  {https://doi.org/10.1103/PhysRevA.105.022405} {\bibfield  {journal} {\bibinfo
   {journal} {Phys. Rev. A}\ }\textbf {\bibinfo {volume} {105}},\ \bibinfo
  {pages} {022405} (\bibinfo {year} {2022}{\natexlab{b}})}\BibitemShut
  {NoStop}%
\bibitem [{\citenamefont {Wang}\ \emph {et~al.}(2023)\citenamefont {Wang},
  \citenamefont {Sager-Smith},\ and\ \citenamefont {Mazziotti}}]{wang2023CQE}%
  \BibitemOpen
  \bibfield  {author} {\bibinfo {author} {\bibfnamefont {Y.}~\bibnamefont
  {Wang}}, \bibinfo {author} {\bibfnamefont {L.~M.}\ \bibnamefont
  {Sager-Smith}},\ and\ \bibinfo {author} {\bibfnamefont {D.~A.}\ \bibnamefont
  {Mazziotti}},\ }\bibfield  {title} {\bibinfo {title} {Quantum simulation of
  bosons with the contracted quantum eigensolver},\ }\href
  {https://doi.org/10.1088/1367-2630/acf9c3} {\bibfield  {journal} {\bibinfo
  {journal} {New J. Phys.}\ }\textbf {\bibinfo {volume} {25}},\ \bibinfo
  {pages} {103005} (\bibinfo {year} {2023})}\BibitemShut {NoStop}%
\bibitem [{\citenamefont {Warren}\ \emph {et~al.}(2024)\citenamefont {Warren},
  \citenamefont {Wang}, \citenamefont {Benavides-Riveros},\ and\ \citenamefont
  {Mazziotti}}]{warren2024}%
  \BibitemOpen
  \bibfield  {author} {\bibinfo {author} {\bibfnamefont {S.}~\bibnamefont
  {Warren}}, \bibinfo {author} {\bibfnamefont {Y.}~\bibnamefont {Wang}},
  \bibinfo {author} {\bibfnamefont {C.~L.}\ \bibnamefont {Benavides-Riveros}},\
  and\ \bibinfo {author} {\bibfnamefont {D.~A.}\ \bibnamefont {Mazziotti}},\
  }\bibfield  {title} {\bibinfo {title} {Exact ansatz of fermion-boson systems
  for a quantum device},\ }\href
  {https://doi.org/10.1103/PhysRevLett.133.080202} {\bibfield  {journal}
  {\bibinfo  {journal} {Phys. Rev. Lett.}\ }\textbf {\bibinfo {volume} {133}},\
  \bibinfo {pages} {080202} (\bibinfo {year} {2024})}\BibitemShut {NoStop}%
\bibitem [{\citenamefont {Mazziotti}(1998)}]{Mazziotti1998}%
  \BibitemOpen
  \bibfield  {author} {\bibinfo {author} {\bibfnamefont {D.~A.}\ \bibnamefont
  {Mazziotti}},\ }\bibfield  {title} {\bibinfo {title} {Contracted
  {S}chr\"odinger equation: Determining quantum energies and two-particle
  density matrices without wave functions},\ }\href
  {https://doi.org/10.1103/PhysRevA.57.4219} {\bibfield  {journal} {\bibinfo
  {journal} {Phys. Rev. A}\ }\textbf {\bibinfo {volume} {57}},\ \bibinfo
  {pages} {4219} (\bibinfo {year} {1998})}\BibitemShut {NoStop}%
\bibitem [{\citenamefont {Mazziotti}(2006)}]{Mazziotti2006}%
  \BibitemOpen
  \bibfield  {author} {\bibinfo {author} {\bibfnamefont {D.~A.}\ \bibnamefont
  {Mazziotti}},\ }\bibfield  {title} {\bibinfo {title} {Anti-hermitian
  contracted {S}chr{\"o}dinger equation: Direct determination of the
  two-electron reduced density matrices of many-electron molecules},\ }\href
  {https://doi.org/10.1103/PhysRevLett.97.143002} {\bibfield  {journal}
  {\bibinfo  {journal} {Phys. Rev. Lett.}\ }\textbf {\bibinfo {volume} {97}},\
  \bibinfo {pages} {143002} (\bibinfo {year} {2006})}\BibitemShut {NoStop}%
\bibitem [{\citenamefont {Nakatsuji}(1976)}]{Nakatsuji1976}%
  \BibitemOpen
  \bibfield  {author} {\bibinfo {author} {\bibfnamefont {H.}~\bibnamefont
  {Nakatsuji}},\ }\bibfield  {title} {\bibinfo {title} {Equation for the direct
  determination of the density matrix},\ }\href
  {https://doi.org/10.1103/PhysRevA.14.41} {\bibfield  {journal} {\bibinfo
  {journal} {Phys. Rev. A}\ }\textbf {\bibinfo {volume} {14}},\ \bibinfo
  {pages} {41} (\bibinfo {year} {1976})}\BibitemShut {NoStop}%
\bibitem [{\citenamefont {Benavides-Riveros}\ \emph {et~al.}(2024)\citenamefont
  {Benavides-Riveros}, \citenamefont {Wang}, \citenamefont {Warren},\ and\
  \citenamefont {Mazziotti}}]{benavides2024}%
  \BibitemOpen
  \bibfield  {author} {\bibinfo {author} {\bibfnamefont {C.~L.}\ \bibnamefont
  {Benavides-Riveros}}, \bibinfo {author} {\bibfnamefont {Y.}~\bibnamefont
  {Wang}}, \bibinfo {author} {\bibfnamefont {S.}~\bibnamefont {Warren}},\ and\
  \bibinfo {author} {\bibfnamefont {D.~A.}\ \bibnamefont {Mazziotti}},\
  }\bibfield  {title} {\bibinfo {title} {Quantum simulation of excited states
  from parallel contracted quantum eigensolvers},\ }\href
  {https://doi.org/10.1088/1367-2630/ad2d1d} {\bibfield  {journal} {\bibinfo
  {journal} {New J. Phys.}\ }\textbf {\bibinfo {volume} {26}},\ \bibinfo
  {pages} {033020} (\bibinfo {year} {2024})}\BibitemShut {NoStop}%
\bibitem [{\citenamefont {Wang}\ and\ \citenamefont
  {Mazziotti}(2023)}]{wang2023}%
  \BibitemOpen
  \bibfield  {author} {\bibinfo {author} {\bibfnamefont {Y.}~\bibnamefont
  {Wang}}\ and\ \bibinfo {author} {\bibfnamefont {D.~A.}\ \bibnamefont
  {Mazziotti}},\ }\bibfield  {title} {\bibinfo {title} {Electronic excited
  states from a variance-based contracted quantum eigensolver},\ }\href
  {https://doi.org/10.1103/PhysRevA.108.022814} {\bibfield  {journal} {\bibinfo
   {journal} {Phys. Rev. A}\ ,\ \bibinfo {pages} {022814}} (\bibinfo {year}
  {2023})}\BibitemShut {NoStop}%
\bibitem [{\citenamefont {Cleve}\ \emph {et~al.}(1998)\citenamefont {Cleve},
  \citenamefont {Ekert}, \citenamefont {Macchiavello},\ and\ \citenamefont
  {Mosca}}]{cleve1998}%
  \BibitemOpen
  \bibfield  {author} {\bibinfo {author} {\bibfnamefont {R.}~\bibnamefont
  {Cleve}}, \bibinfo {author} {\bibfnamefont {A.}~\bibnamefont {Ekert}},
  \bibinfo {author} {\bibfnamefont {C.}~\bibnamefont {Macchiavello}},\ and\
  \bibinfo {author} {\bibfnamefont {M.}~\bibnamefont {Mosca}},\ }\bibfield
  {title} {\bibinfo {title} {Quantum algorithms revisited},\ }\href
  {https://doi.org/10.1098/rspa.1998.0164} {\bibfield  {journal} {\bibinfo
  {journal} {Proc. R. Soc. London A}\ }\textbf {\bibinfo {volume} {454}},\
  \bibinfo {pages} {339} (\bibinfo {year} {1998})}\BibitemShut {NoStop}%
\bibitem [{\citenamefont {Alcoba}\ \emph {et~al.}(2024)\citenamefont {Alcoba},
  \citenamefont {Lain}, \citenamefont {Torre}, \citenamefont {Ayala},
  \citenamefont {O{\~n}a}, \citenamefont {Massaccesi}, \citenamefont
  {Peralta},\ and\ \citenamefont {Melo}}]{alcoba2024}%
  \BibitemOpen
  \bibfield  {author} {\bibinfo {author} {\bibfnamefont {D.~R.}\ \bibnamefont
  {Alcoba}}, \bibinfo {author} {\bibfnamefont {L.}~\bibnamefont {Lain}},
  \bibinfo {author} {\bibfnamefont {A.}~\bibnamefont {Torre}}, \bibinfo
  {author} {\bibfnamefont {T.~R.}\ \bibnamefont {Ayala}}, \bibinfo {author}
  {\bibfnamefont {O.~B.}\ \bibnamefont {O{\~n}a}}, \bibinfo {author}
  {\bibfnamefont {G.~E.}\ \bibnamefont {Massaccesi}}, \bibinfo {author}
  {\bibfnamefont {J.~E.}\ \bibnamefont {Peralta}},\ and\ \bibinfo {author}
  {\bibfnamefont {J.~I.}\ \bibnamefont {Melo}},\ }\bibfield  {title} {\bibinfo
  {title} {Generalized spin in the variance-based wave function optimization
  method within the doubly occupied configuration interaction framework},\
  }\bibfield  {journal} {\bibinfo  {journal} {J. Phys. Chem. A}\ }\href
  {https://doi.org/10.1021/acs.jpca.4c02742} {10.1021/acs.jpca.4c02742}
  (\bibinfo {year} {2024})\BibitemShut {NoStop}%
\bibitem [{\citenamefont {Nakanishi}\ \emph {et~al.}(2019)\citenamefont
  {Nakanishi}, \citenamefont {Mitarai},\ and\ \citenamefont
  {Fujii}}]{nakanishi2019}%
  \BibitemOpen
  \bibfield  {author} {\bibinfo {author} {\bibfnamefont {K.~M.}\ \bibnamefont
  {Nakanishi}}, \bibinfo {author} {\bibfnamefont {K.}~\bibnamefont {Mitarai}},\
  and\ \bibinfo {author} {\bibfnamefont {K.}~\bibnamefont {Fujii}},\ }\bibfield
   {title} {\bibinfo {title} {Subspace-search variational quantum eigensolver
  for excited states},\ }\href@noop {} {\bibfield  {journal} {\bibinfo
  {journal} {Phys. Rev. Res.}\ }\textbf {\bibinfo {volume} {1}},\ \bibinfo
  {pages} {033062} (\bibinfo {year} {2019})}\BibitemShut {NoStop}%
\bibitem [{\citenamefont {Lischka}\ \emph {et~al.}(2011)\citenamefont
  {Lischka}, \citenamefont {M{\"u}ller}, \citenamefont {Szalay}, \citenamefont
  {Shavitt}, \citenamefont {Pitzer},\ and\ \citenamefont
  {Shepard}}]{lischka2011}%
  \BibitemOpen
  \bibfield  {author} {\bibinfo {author} {\bibfnamefont {H.}~\bibnamefont
  {Lischka}}, \bibinfo {author} {\bibfnamefont {T.}~\bibnamefont {M{\"u}ller}},
  \bibinfo {author} {\bibfnamefont {P.~G.}\ \bibnamefont {Szalay}}, \bibinfo
  {author} {\bibfnamefont {I.}~\bibnamefont {Shavitt}}, \bibinfo {author}
  {\bibfnamefont {R.~M.}\ \bibnamefont {Pitzer}},\ and\ \bibinfo {author}
  {\bibfnamefont {R.}~\bibnamefont {Shepard}},\ }\bibfield  {title} {\bibinfo
  {title} {Columbus—a program system for advanced multireference theory
  calculations},\ }\href@noop {} {\bibfield  {journal} {\bibinfo  {journal}
  {Wiley Interdiscip. Rev. Comput. Mol. Sci.}\ }\textbf {\bibinfo {volume}
  {1}},\ \bibinfo {pages} {191} (\bibinfo {year} {2011})}\BibitemShut {NoStop}%
\bibitem [{\citenamefont {Lischka}\ \emph {et~al.}(2020)\citenamefont
  {Lischka}, \citenamefont {Shepard}, \citenamefont {M{\"u}ller}, \citenamefont
  {Szalay}, \citenamefont {Pitzer}, \citenamefont {Aquino}, \citenamefont
  {Ara{\'u}jo~do Nascimento}, \citenamefont {Barbatti}, \citenamefont
  {Belcher}, \citenamefont {Blaudeau} \emph {et~al.}}]{lischka2020}%
  \BibitemOpen
  \bibfield  {author} {\bibinfo {author} {\bibfnamefont {H.}~\bibnamefont
  {Lischka}}, \bibinfo {author} {\bibfnamefont {R.}~\bibnamefont {Shepard}},
  \bibinfo {author} {\bibfnamefont {T.}~\bibnamefont {M{\"u}ller}}, \bibinfo
  {author} {\bibfnamefont {P.~G.}\ \bibnamefont {Szalay}}, \bibinfo {author}
  {\bibfnamefont {R.~M.}\ \bibnamefont {Pitzer}}, \bibinfo {author}
  {\bibfnamefont {A.~J.}\ \bibnamefont {Aquino}}, \bibinfo {author}
  {\bibfnamefont {M.~M.}\ \bibnamefont {Ara{\'u}jo~do Nascimento}}, \bibinfo
  {author} {\bibfnamefont {M.}~\bibnamefont {Barbatti}}, \bibinfo {author}
  {\bibfnamefont {L.~T.}\ \bibnamefont {Belcher}}, \bibinfo {author}
  {\bibfnamefont {J.-P.}\ \bibnamefont {Blaudeau}}, \emph {et~al.},\ }\bibfield
   {title} {\bibinfo {title} {The generality of the guga mrci approach in
  columbus for treating complex quantum chemistry},\ }\href@noop {} {\bibfield
  {journal} {\bibinfo  {journal} {J. Chem. Phys.}\ }\textbf {\bibinfo {volume}
  {152}} (\bibinfo {year} {2020})}\BibitemShut {NoStop}%
\bibitem [{\citenamefont {Manaa}\ and\ \citenamefont
  {Yarkony}(1993)}]{manaa1993}%
  \BibitemOpen
  \bibfield  {author} {\bibinfo {author} {\bibfnamefont {M.~R.}\ \bibnamefont
  {Manaa}}\ and\ \bibinfo {author} {\bibfnamefont {D.~R.}\ \bibnamefont
  {Yarkony}},\ }\bibfield  {title} {\bibinfo {title} {On the intersection of
  two potential energy surfaces of the same symmetry. systematic
  characterization using a lagrange multiplier constrained procedure},\
  }\href@noop {} {\bibfield  {journal} {\bibinfo  {journal} {J. Chem. Phys.}\
  }\textbf {\bibinfo {volume} {99}},\ \bibinfo {pages} {5251} (\bibinfo {year}
  {1993})}\BibitemShut {NoStop}%
\bibitem [{\citenamefont {Cu{\'e}llar-Zuquin}\ \emph
  {et~al.}(2024)\citenamefont {Cu{\'e}llar-Zuquin}, \citenamefont {Pepino},
  \citenamefont {Galv{\'a}n}, \citenamefont {Rivalta}, \citenamefont
  {Aquilante}, \citenamefont {Garavelli}, \citenamefont {Lindh},\ and\
  \citenamefont {Segarra-Mart{\'\i}}}]{cuellar2023}%
  \BibitemOpen
  \bibfield  {author} {\bibinfo {author} {\bibfnamefont {J.}~\bibnamefont
  {Cu{\'e}llar-Zuquin}}, \bibinfo {author} {\bibfnamefont {A.~J.}\ \bibnamefont
  {Pepino}}, \bibinfo {author} {\bibfnamefont {I.~F.}\ \bibnamefont
  {Galv{\'a}n}}, \bibinfo {author} {\bibfnamefont {I.}~\bibnamefont {Rivalta}},
  \bibinfo {author} {\bibfnamefont {F.}~\bibnamefont {Aquilante}}, \bibinfo
  {author} {\bibfnamefont {M.}~\bibnamefont {Garavelli}}, \bibinfo {author}
  {\bibfnamefont {R.}~\bibnamefont {Lindh}},\ and\ \bibinfo {author}
  {\bibfnamefont {J.}~\bibnamefont {Segarra-Mart{\'\i}}},\ }\bibfield  {title}
  {\bibinfo {title} {Characterising conical intersections in {DNA/RNA}
  nucleobases with multiconfigurational wave functions of varying active space
  size},\ }\href {https://doi.org/10.1021/acs.jctc.3c00577} {\bibfield
  {journal} {\bibinfo  {journal} {J. Chem. Theory Comput.}\ }\textbf {\bibinfo
  {volume} {19}},\ \bibinfo {pages} {8258} (\bibinfo {year}
  {2024})}\BibitemShut {NoStop}%
\bibitem [{\citenamefont {{Qiskit contributors}}(2024)}]{Qiskit}%
  \BibitemOpen
  \bibfield  {author} {\bibinfo {author} {\bibnamefont {{Qiskit
  contributors}}},\ }\href@noop {} {\bibinfo {title} {Qiskit: An open-source
  framework for quantum computing}} (\bibinfo {year} {2024})\BibitemShut
  {NoStop}%
\bibitem [{\citenamefont {Jordan}\ and\ \citenamefont
  {Wigner}(1928)}]{Jordan1928}%
  \BibitemOpen
  \bibfield  {author} {\bibinfo {author} {\bibfnamefont {P.}~\bibnamefont
  {Jordan}}\ and\ \bibinfo {author} {\bibfnamefont {E.}~\bibnamefont
  {Wigner}},\ }\bibfield  {title} {\bibinfo {title} {Über das paulische
  Äquivalenzverbot},\ }\href {https://doi.org/10.1007/bf01331938} {\bibfield
  {journal} {\bibinfo  {journal} {Z. Physik}\ }\textbf {\bibinfo {volume}
  {47}},\ \bibinfo {pages} {631–651} (\bibinfo {year} {1928})}\BibitemShut
  {NoStop}%
\bibitem [{\citenamefont {IBM-Quantum}(2024)}]{ibm_quantum}%
  \BibitemOpen
  \bibfield  {author} {\bibinfo {author} {\bibnamefont {IBM-Quantum}},\
  }\href@noop {} {\bibinfo {title} {https://quantum-computing.ibm.com/}}
  (\bibinfo {year} {2024})\BibitemShut {NoStop}%
\bibitem [{\citenamefont {Smart}\ and\ \citenamefont
  {Mazziotti}(2024)}]{smart2024}%
  \BibitemOpen
  \bibfield  {author} {\bibinfo {author} {\bibfnamefont {S.~E.}\ \bibnamefont
  {Smart}}\ and\ \bibinfo {author} {\bibfnamefont {D.~A.}\ \bibnamefont
  {Mazziotti}},\ }\bibfield  {title} {\bibinfo {title} {Verifiably exact
  solution of the electronic schr{\"o}dinger equation on quantum devices},\
  }\href {https://doi.org/10.1103/PhysRevA.109.022802} {\bibfield  {journal}
  {\bibinfo  {journal} {Phys. Rev. A}\ }\textbf {\bibinfo {volume} {109}},\
  \bibinfo {pages} {022802} (\bibinfo {year} {2024})}\BibitemShut {NoStop}%
\bibitem [{\citenamefont {Mazziotti}(2007)}]{Mazziotti.2007}%
  \BibitemOpen
  \bibfield  {author} {\bibinfo {author} {\bibfnamefont {D.~A.}\ \bibnamefont
  {Mazziotti}},\ }\bibfield  {title} {\bibinfo {title} {{Anti-Hermitian part of
  the contracted Schrödinger equation for the direct calculation of
  two-electron reduced density matrices}},\ }\href
  {https://doi.org/10.1103/physreva.75.022505} {\bibfield  {journal} {\bibinfo
  {journal} {Phys. Rev. A}\ }\textbf {\bibinfo {volume} {75}},\ \bibinfo
  {pages} {022505} (\bibinfo {year} {2007})}\BibitemShut {NoStop}%
\bibitem [{\citenamefont {Mazziotti}(2004)}]{Mazziotti.2004}%
  \BibitemOpen
  \bibfield  {author} {\bibinfo {author} {\bibfnamefont {D.~A.}\ \bibnamefont
  {Mazziotti}},\ }\bibfield  {title} {\bibinfo {title} {{Exactness of wave
  functions from two-body exponential transformations in many-body quantum
  theory}},\ }\href {https://doi.org/10.1103/physreva.69.012507} {\bibfield
  {journal} {\bibinfo  {journal} {Physical Review A}\ }\textbf {\bibinfo
  {volume} {69}},\ \bibinfo {pages} {012507} (\bibinfo {year}
  {2004})}\BibitemShut {NoStop}%
\bibitem [{\citenamefont {Hoffmann}\ and\ \citenamefont
  {Simons}(1988)}]{Hoffmann1988}%
  \BibitemOpen
  \bibfield  {author} {\bibinfo {author} {\bibfnamefont {M.~R.}\ \bibnamefont
  {Hoffmann}}\ and\ \bibinfo {author} {\bibfnamefont {J.}~\bibnamefont
  {Simons}},\ }\bibfield  {title} {\bibinfo {title} {A unitary
  multiconfigurational coupled-cluster method: Theory and applications},\
  }\href {https://doi.org/10.1063/1.454125} {\bibfield  {journal} {\bibinfo
  {journal} {J. Chem. Phys.}\ }\textbf {\bibinfo {volume} {88}},\ \bibinfo
  {pages} {993–1002} (\bibinfo {year} {1988})}\BibitemShut {NoStop}%
\bibitem [{\citenamefont {Romero}\ \emph {et~al.}(2018)\citenamefont {Romero},
  \citenamefont {Babbush}, \citenamefont {McClean}, \citenamefont {Hempel},
  \citenamefont {Love},\ and\ \citenamefont {Aspuru-Guzik}}]{romero2018}%
  \BibitemOpen
  \bibfield  {author} {\bibinfo {author} {\bibfnamefont {J.}~\bibnamefont
  {Romero}}, \bibinfo {author} {\bibfnamefont {R.}~\bibnamefont {Babbush}},
  \bibinfo {author} {\bibfnamefont {J.~R.}\ \bibnamefont {McClean}}, \bibinfo
  {author} {\bibfnamefont {C.}~\bibnamefont {Hempel}}, \bibinfo {author}
  {\bibfnamefont {P.~J.}\ \bibnamefont {Love}},\ and\ \bibinfo {author}
  {\bibfnamefont {A.}~\bibnamefont {Aspuru-Guzik}},\ }\bibfield  {title}
  {\bibinfo {title} {Strategies for quantum computing molecular energies using
  the unitary coupled cluster ansatz},\ }\href
  {https://doi.org/10.1088/2058-9565/aad3e4} {\bibfield  {journal} {\bibinfo
  {journal} {Quantum Sci. Technol.}\ }\textbf {\bibinfo {volume} {4}},\
  \bibinfo {pages} {014008} (\bibinfo {year} {2018})}\BibitemShut {NoStop}%
\bibitem [{\citenamefont {Bonet-Monroig}\ \emph {et~al.}(2020)\citenamefont
  {Bonet-Monroig}, \citenamefont {Babbush},\ and\ \citenamefont
  {O’Brien}}]{bonet2020}%
  \BibitemOpen
  \bibfield  {author} {\bibinfo {author} {\bibfnamefont {X.}~\bibnamefont
  {Bonet-Monroig}}, \bibinfo {author} {\bibfnamefont {R.}~\bibnamefont
  {Babbush}},\ and\ \bibinfo {author} {\bibfnamefont {T.~E.}\ \bibnamefont
  {O’Brien}},\ }\bibfield  {title} {\bibinfo {title} {Nearly optimal
  measurement scheduling for partial tomography of quantum states},\ }\href
  {https://doi.org/10.1103/PhysRevX.10.031064} {\bibfield  {journal} {\bibinfo
  {journal} {Phys. Rev. X}\ }\textbf {\bibinfo {volume} {10}},\ \bibinfo
  {pages} {031064} (\bibinfo {year} {2020})}\BibitemShut {NoStop}%
\bibitem [{\citenamefont {Wang}\ \emph {et~al.}(2024)\citenamefont {Wang},
  \citenamefont {Avdic},\ and\ \citenamefont {Mazziotti}}]{wang20242}%
  \BibitemOpen
  \bibfield  {author} {\bibinfo {author} {\bibfnamefont {Y.}~\bibnamefont
  {Wang}}, \bibinfo {author} {\bibfnamefont {I.}~\bibnamefont {Avdic}},\ and\
  \bibinfo {author} {\bibfnamefont {D.~A.}\ \bibnamefont {Mazziotti}},\
  }\bibfield  {title} {\bibinfo {title} {Shadow ansatz for the many-fermion
  wave function in scalable molecular simulations on quantum computing
  devices},\ }\bibfield  {journal} {\bibinfo  {journal} {arXiv preprint
  arXiv:2408.11026}\ }\href {https://doi.org/10.48550/arXiv.2408.11026}
  {10.48550/arXiv.2408.11026} (\bibinfo {year} {2024})\BibitemShut {NoStop}%
\bibitem [{\citenamefont {Spall}(1998)}]{spall1998}%
  \BibitemOpen
  \bibfield  {author} {\bibinfo {author} {\bibfnamefont {J.~C.}\ \bibnamefont
  {Spall}},\ }\bibfield  {title} {\bibinfo {title} {Implementation of the
  simultaneous perturbation algorithm for stochastic optimization},\ }\href
  {https://doi.org/10.1109/7.705889} {\bibfield  {journal} {\bibinfo  {journal}
  {IEEE Trans. Aerosp. Electron. Syst.}\ }\textbf {\bibinfo {volume} {34}},\
  \bibinfo {pages} {817} (\bibinfo {year} {1998})}\BibitemShut {NoStop}%
\end{thebibliography}%
\end{document}